\documentclass[english,journal=jcisd8,manuscript=article,layout=onecolumn]{achemso}
\usepackage[LGR,T1]{fontenc}
\usepackage{textcomp}
\usepackage[latin9]{inputenc}
\usepackage{array}
\usepackage{refstyle}
\usepackage{booktabs}
\usepackage{units}
\usepackage{amsmath}
\usepackage{amssymb}
\usepackage{stackrel}
\usepackage{graphicx}
\usepackage{geometry}
\makeatletter

\author{Steven Ayoub{*}}

\affiliation{Department of Chemistry and Biochemistry, California State University,
Northridge, Northridge, CA 91330}

\alsoaffiliation{Department of Pharmaceutical Sciences, University of California,
Irvine CA 92697}

\altaffiliation{Contributed equally to this work}

\author{Michael Barton{*}}

\affiliation{Department of Physics and Astronomy, California State University,
Northridge, Northridge, CA 91330}

\altaffiliation{Contributed equally to this work}

\author{David A. Case}

\affiliation{Department of Chemistry and Chemical Biology, Rutgers University,
Piscataway, NJ}

\author{Tyler Luchko}

\affiliation{Department of Physics and Astronomy, California State University,
Northridge, Northridge, CA 91330}

\email{tluchko@csun.edu}

\title{Automated Workflow for Absolute Binding Free Energy Calculations
with Implicit Solvent and Double Decoupling}

\AtBeginDocument{\providecommand\figref[1]{\ref{fig:#1}}}
\AtBeginDocument{\providecommand\tabref[1]{\ref{tab:#1}}}
\AtBeginDocument{\providecommand\Figref[1]{\ref{Fig:#1}}}
\AtBeginDocument{\providecommand\secref[1]{\ref{sec:#1}}}
\AtBeginDocument{\providecommand\Tabref[1]{\ref{Tab:#1}}}
\DeclareRobustCommand{\greektext}{%
  \fontencoding{LGR}\selectfont\def\encodingdefault{LGR}}
\DeclareRobustCommand{\textgreek}[1]{\leavevmode{\greektext #1}}

\newcommand{\lyxmathsym}[1]{\ifmmode\begingroup\def\b@ld{bold}
  \text{\ifx\math@version\b@ld\bfseries\fi#1}\endgroup\else#1\fi}

\providecommand{\tabularnewline}{\\}
\RS@ifundefined{subsecref}
  {\newref{subsec}{name = \RSsectxt}}
  {}
\RS@ifundefined{thmref}
  {\def\RSthmtxt{theorem~}\newref{thm}{name = \RSthmtxt}}
  {}
\RS@ifundefined{lemref}
  {\def\RSlemtxt{lemma~}\newref{lem}{name = \RSlemtxt}}
  {}

\providecommand*{\code}[1]{\texttt{#1}}

\SectionNumbersOn

\renewcommand{\tabref}{\Tabref}
\renewcommand{\figref}{\Figref}

\makeatother

\usepackage{babel}
\begin{document}
\begin{abstract}
Accurate absolute binding free energy (ABFE) calculations can reduce
the time and cost of identifying drug candidates from a diverse pool
of molecules that may have been overlooked experimentally. These calculations
typically employ explicit solvents; however these models can be computationally
demanding and challenging to implement due to the difficulties in
sampling waters and managing changes in net charge. To address these
challenges, we introduce an automated parallel Python workflow that
adopts the double decoupling method, incorporating conformational
restraints and pairing it with the implicit generalized Born (GB)
solvation model. This approach enhances convergence, reduces computational
costs and avoids the technical issues associated with explicit solvents.
We applied this workflow to a series of 93 host-guest complexes from
the TapRoom database. When pooling all systems, the GB(OBC) model
correlated well with experiment (R\texttwosuperior{} \ensuremath{\approx}
0.86); however, this global metric obscured much weaker correlations
observed within individual hosts (R\texttwosuperior{} = 0.3--0.8).
Systematic errors associated with charged functional groups (notably
ammonium and carboxylates) were also evident, resulting in root-mean-squared
errors (RMSEs) greater than 6.12 kcal/mol across all models. Although
the GB models did not achieve reliable accuracy across all systems,
they may be practical when all ligands contain the same functional
groups. Furthermore, a linear correction based on these functional
groups reduced RMSE values to within \textasciitilde 1 kcal/mol of
experiment, and our error analysis suggests specific changes for future
GB models to improve accuracy. Overall, the Python workflow demonstrates
promise for fast, reliable absolute binding free energy calculations
by leveraging the sampling efficiency of GB in a fully automated application.
\end{abstract}

\section{Introduction}

Binding free energy (BFE) calculations are used in drug discovery
to rank potential candidate compounds by the estimating the binding
affinity of a ligand to a target host molecule, reducing the cost
of expensive, physical methods in drug testing. \cite{dengComputationsStandardBinding2009a,ganesan2017molecular,salo-ahen2021molecular}
Relative binding free energy (RBFE) calculations, which predict the
difference in BFE between ligand pairs, play a significant role in
the lead optimization in the later states of drug discovery, such
as making small changes to a lead candidate to create higher affinity.
\cite{salo-ahen2021molecular,hollingsworth2018molecular,ganesan2017molecular,muddanaSAMPL4HostGuest2014b,heinzelmannAutomatedDockingRefinement2020a,schindlerLargeScaleAssessmentBinding2020a}
However, RBFE calculations generally require structural and chemical
similarity between molecules, which limits its capability of exploring
chemical space to find novel lead compounds early in the drug discovery
stage. \cite{courniaRigorousFreeEnergy2020b}

Absolute binding free energy (ABFE) calculations predict the BFE for
individual molecules, allowing screening of diverse chemical compounds
without a common scaffold \cite{courniaRigorousFreeEnergy2020b}.
Several approaches exist, including end-point methods such as molecular
mechanics with Poisson-Boltzmann (MM/PBSA) \cite{hou2011assessing,miller2012mmpbsapy,srinivasan1998continuum}
and linear interaction energy (LIE) \cite{raqvist1994anew,lee1992calculations},
pathway-based techniques like attach-pull-release \cite{henriksen2015computational,heinzelmann2017attachpullrelease},
or alchemical methods like single decoupling\cite{kilburg2018assessment,sakae2020absolute}.
In recent years, the double decoupling method (DDM) has seen renewed
interest due to its theoretically rigorous approach and broad applicability\cite{boresch2003absolute,gilsonStatisticalthermodynamicBasisComputation1997b}.
DDM uses the bound and unbound end states and a series of intermediate
states where the ligand is decoupled from both the environment and
its interactions with its host. Starting from the bound-state, the
ligand is decoupled from both the solvent and receptor in a series
of steps by scaling its electrostatics and van der Waals (VDW) interactions
to zero, while orientational and distance restraints keep the ligand
in the binding pocket. \cite{boresch2003absolute} Then, the ligand
is recoupled to the solvent in isolation, which is the unbound state.
This allows the receptor and ligand to be brought together without
sampling complicated interactions during the binding process.

However, the use of explicit water molecules causes DDM to be computationally
demanding because all the water atoms must be simulated, and also
because friction with the water can slow conformational change and
some water molecules may have long residency times \cite{durrantMolecularDynamicsSimulations2011a,anandakrishnan_speed_2015}.
The large number of degrees of freedom contributed by the water also
makes it difficult to overcome these challenges with enhanced sampling
methods \cite{khalak2021alchemical,aldeghi2017predictions}. Commonly
used methods, such as adaptive force bias \cite{darve2008adaptive},
umbrella sampling \cite{patey1975amonte}, steered MD \cite{griebel_steered_1999},
metadynamics \cite{laio_escaping_2002}, require advanced knowledge
of the critical collective variables and may not improve conformational
sampling in general. Methods that do promote general conformational
sampling, such as replica exchange molecular dynamics (REMD) \cite{sugita1999replicaexchange},
are often rendered impractical by the large number of degrees of freedom
that explicit water contribute. This may lead to inadequate conformational
sampling, especially of the bound and unbound end-states, and produce
inaccurate results \cite{durrantMolecularDynamicsSimulations2011a,onuchic2004theoryof}.

In addition, explicit solvent contributes to the complexity of setting
up and running DDM calculations \cite{deng_comparing_2018}. Because
explicit solvent calculations are generally run with periodic boundaries,
corrections are required to treat artifacts due to finite size, periodicity
and changes in net charge. \cite{ohlknecht2020correcting,rocklin2013calculating}
Also, decoupling of VDW interactions between the water and ligand
results in steric overlaps between solvent and solute, creating enormous
potential energies and numerical instabilities. This has led to the
use of soft-core potentials in order to minimize or resolve the instabilities
observed during the decoupling of the VDW parameters, a step that
can be complex and takes care to implement. \cite{leeAlchemicalBindingFree2020a,pohorille2010goodpractices,steinbrecher2007nonlinear,beutler_avoiding_1994}

The generalized Born (GB) implicit solvent model \cite{still1990semianalytical}
is an attractive alternative to explicit solvents because of its computational
speed and solute conformational sampling efficiency \cite{anandakrishnan_speed_2015}.
Solute conformational sampling may be further aided by implicit solvents
because the solvent degrees of freedom (DOF) are integrated out, allowing
many fewer replicas to be used than with explicit solvent. However,
GB replaces the atomistic detail of explicit solvent with an approximate
dielectric continuum, which affects its accuracy and has rarely been
used with DDM or other ABFE methods with similarly rigorous theoretical
foundations \cite{sakae2020absolute,cruz2020combining,kilburg2018assessment,gallicchio2010binding,setiadi2024tuningpotential}.

To reduce the computational cost of DDM ABFE calculations and improve
reproducibility, we have created an automated workflow based on the
DDM thermodynamic cycle, which we modified to include conformational
restraints (which here are harmonic distance restraints between atoms
separated by less that 6 $\textrm{\AA}$), and to employ the GB implicit
solvent model. This approach overcomes several challenges with explicit
solvent DDM calculations. Computational costs are much lower than
explicit solvent DDM calculations to achieve the same numerical precision,
due directly to the use of GB and the ability to use temperature REMD
(TREMD) to sample end-state conformations. Soft-core Lennard-Jones
models are unnecessary, as steric overlaps are eliminated when the
ligand is in the binding site and decoupled from the receptor and
solvent. This is a consequence of employing GB, Boresch restraints
and conformational restraints. Conformational restraints also limit
accessible conformational space in the intermediate states, which
reduces the sampling required in these states. Our implementation
of this approach aids reproducibility and reduces complexity for the
user by automating the process, enhanced sampling for end states and
removing time correlations during analysis. In what follows, we demonstrate
the reproducibility and numerical precision of the workflow and assess
the accuracy of several commonly used GB models in the AmberTools
molecular modeling suite \cite{case2023ambertools} by calculating
the binding affinity of 93 host-guest molecules from the Taproom database
\cite{slochowerTaproom2022}. While none of the GB models we tested
achieved the accuracy needed for consistently reliable predictions
across all systems, they may still be useful in specific applications
where the ligands are composed of common functional groups. An analysis
of the errors suggest changes that might be incorporated into future
GB or other implicit solvent models.

\section{Methods\label{sec:methods}}

\subsection{Conformationally restrained DDM\label{subsec:Our-Alchemical-Pathway}}

\begin{figure*}
\begin{centering}
\includegraphics[width=5in]{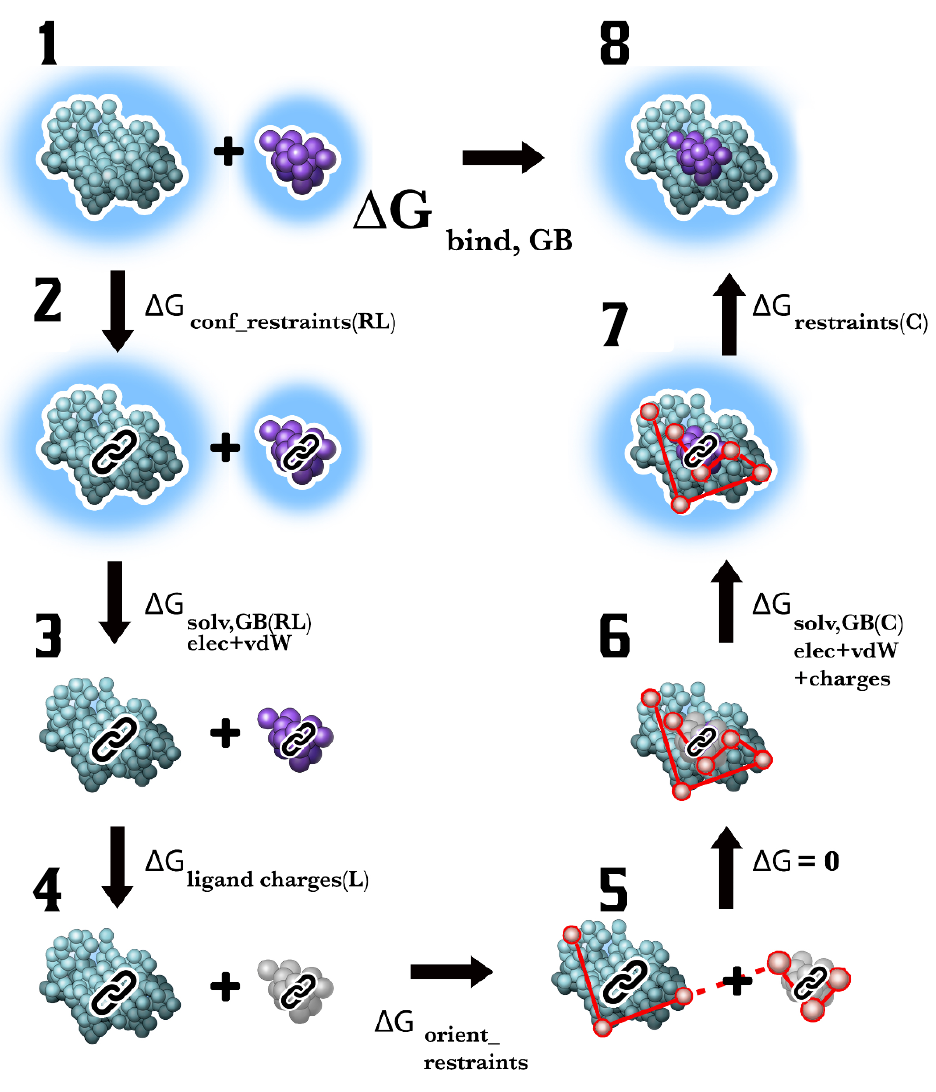}
\par\end{centering}
\caption{Major states in our binding free energy thermodynamic cycle. The free
energy change of binding in GB solvent is the free energy difference
between states 1 and 8. Conformational restraints (padlock) indicates
the presence of harmonic distance restraints between each atom separated
by $<6\,\text{Å}$. The filled purple circle indicates that ligand
charges are active, while the unfilled circle indicates charges have
been set to 0. The light blue background is representation of GB solvent.
Orientational restraints (red dotted line) are applied in states 5
and 6\cite{boresch2003absolute}. State 6 is where the ligand is decoupled
from the receptor but is orientationally restrained relative to receptor.
State 5 is equivalent to state 6 and not simulated.\label{fig:OurCycle}}
\end{figure*}

Our thermodynamic cycle (\figref{OurCycle}), connects the bound and
unbound end-states using DDM-like procedure, modified to employ conformational
restraints to exploit the benefits implicit solvent models, as follows.
\begin{description}
\item [{State~1}] Simulation of the unbound end state.
\item [{State~2}] Adding conformational restraints to the host and ligand.
\item [{State~3}] MD simulations of the receptor and ligand with conformational
restraints in vacuum (no GB).
\item [{State~4}] MD simulations in vacuum, with conformational restraints,
and ligand partial charges set to zero.
\item [{State~5}] Boresch orientational restraints are introduced analytically.
\cite{boresch2003absolute}
\item [{State~6-LJ}] MD simulations in vacuum, with conformational and
orientation restraints, and ligand partial charges set to zero. There
are no interactions between the receptor and ligand.
\item [{State~6-GB}] MD simulations in GB solvent, with conformational
and orientation restraints, and ligand partial chargers set to zero.
The receptor and ligand fully interact. 
\item [{State~7}] MD simulations in GB solvent, with conformational and
orientation restraints, and full ligand partial charges.
\item [{State~7-restraints}] Conformational and orientational restraints
are removed through a series of MD simulations.
\item [{State~8}] Simulation of the bound end state, with only flat-bottom
harmonic restraints between the receptor and ligand.
\end{description}
While these are the major states in the cycle, restraints, charges
and solvent are introduced and removed in intermediate states that
may vary in number, depending on the system. The binding free energy
may be then calculated as
\begin{align*}
\Delta G_{\text{bind}} & =\Delta G_{1,8}\\
 & =\Delta G_{1,2}+\Delta G_{2,3}+\Delta G_{3,4}\\
 & \phantom{=}+\Delta G_{4,5}+\Delta G_{5,6}+\Delta G_{7,8}.
\end{align*}

As in other end-state methods, exhaustive conformational sampling
of the bound and unbound end-states (states 1 and 8) is required,
but a particular bound conformation is selected as the path of calculating
the binding free energy. The free energy cost of selecting one conformation
is accounted for in $\Delta G_{1,2}$ and $\Delta G_{7,8}$. While
there is a computational cost of simulations to add and remove the
conformational restraints, these restraints restrict the motion of
the ligand and receptor in states 3 to 6. This limits the accessible
conformational space in intermediate states and, when an implicit
solvent is used, removes the need for soft-core Lennard-Jones interactions.

\begin{table*}
\begin{centering}
\begin{tabular}{llllllllllll}
\toprule 
State & 1 & 2 & 3 & 4 & 5 & 6 & 6-LJ & 6-Coul & 6-GB & 7 & 8\tabularnewline
\midrule
GB Solvent & 1 & 1 & S & 0 & -- & 0 & 0 & 0 & S & 1 & 1\tabularnewline
Conformational Restraints & 0 & 1 & 1 & 1 & -- & 1 & 1 & 1 & 1 & 1 & 0\tabularnewline
Solvent Interactions & 1 & 1 & 0 & 0 & -- & 0 & 0 & 0 & 0 & 1 & 1\tabularnewline
Molecular Interactions & 1 & 1 & 0 & 0 & -- & 0 & 1 & 1 & 1 & 1 & 1\tabularnewline
Ligand Charge & 1 & 1 & 1 & S & -- & 0 & 0 & S & 1 & 1 & 1\tabularnewline
Orientational Restraints & 0 & 0 & 0 & 0 & 1 & 1 & 1 & 1 & 1 & 1 & 0\tabularnewline
Ligand+Receptor or Complex & LR & LR & LR & LR & LR & C & C & C & C & C & C\tabularnewline
\bottomrule
\end{tabular}
\par\end{centering}
\caption{Simulation settings for each state of in \figref{OurCycle}. 0: off,
1: on, S: scaled, LR: ligand-receptor, C: complex, --: not applicable.
\label{tab:alchemical-steps}}
\end{table*}

\subsection{Implementation Details of \protect\code{ISDDM.py} Workflow \label{sec:Implementation-Details}}

\begin{figure*}
\includegraphics{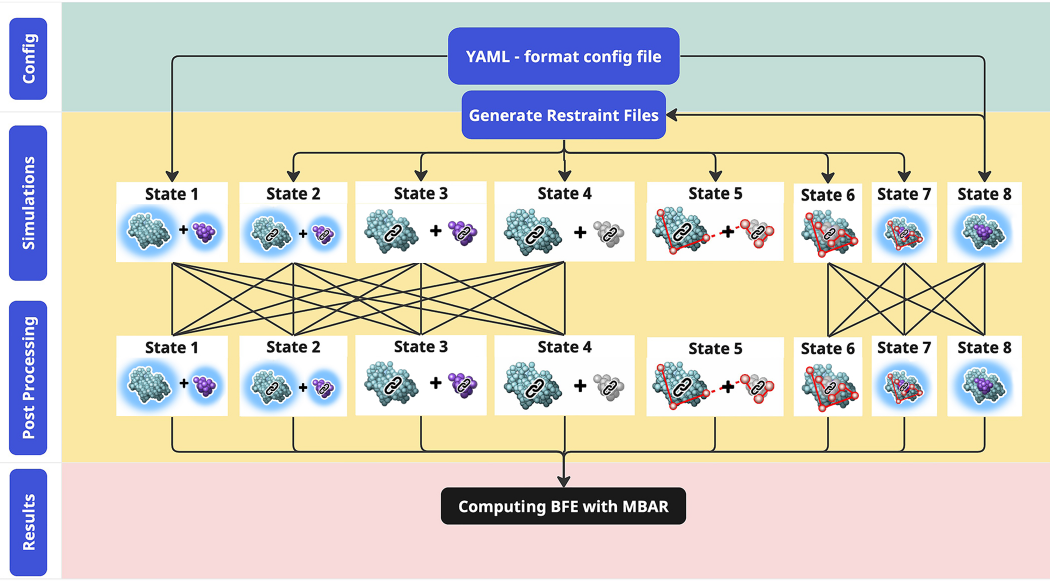}\caption{Automated workflow pipeline for \protect\code{ISDDM.py}. Users provide
a force field parameter file for the ligand and receptor, and either
a set of initial coordinates for the bound state or trajectory files
of conformational samples of the bound and unbound end-states. The
binding free energy and estimated errors are output at the end of
the workflow.\label{fig:Automated-workflow}}
\end{figure*}

\code{ISDDM.py} is our practical implementation of this thermodynamic
cycle using the Amber and AmberTools molecular modeling suite \cite{case2023ambertools}.
Using an initial structure of the receptor-ligand complex parameterized
for Amber, it automates preparing, running and post-processing all
the MD simulations for calculating the free energies (see \figref{Automated-workflow}).
Details of the cycle are controlled through a YAML format configure
file and two MDIN files, one for the MD settings of the end-states
and another for the intermediate states. Each major state is implemented
as follows (\tabref{alchemical-steps}):
\begin{description}
\item [{State~1}] (optional) TREMD or standard MD simulations of the unbound
receptor and ligand.
\item [{States~1-2}] Conformational restraints are introduced through
a series of MD simulations. The value of the restraint force constant
for each simulation is specified by the user.
\item [{States~2-3}] GB dielectric constant is reduced in a series of
conformationally restrained simulations of the receptor and ligand
until a vacuum is reached (\code{igb=6}).
\item [{States~3-4}] Partial charges are scaled to zero using \code{parmed}
in a series of conformationally restrained simulations of the ligand
in a vacuum.
\item [{States~4-6}] Analytic introduction of orientational restraints
(no simulations).
\item [{States~6-LJ}] Simulation of the conformationally and orientationally
restrained receptor and ligand in vacuum with only LJ interactions
between the receptor and ligand.
\item [{States~6-GB}] A series of simulations reintroducing the GB solvation
model for the conformationally and orientationally restrained receptor
and ligand in vacuum with LJ interactions and ligand partial charges
set to 0. 
\item [{States~6-Coul}] A series of simulations reinstating ligand partial-charges
for the conformationally and orientationally restrained receptor and
ligand in GB solvation with LJ interactions.
\item [{States~6-7}] Receptor and ligand are simulated together with orientational
restraints. LJ interactions are turned on between ligand and receptor,
ligand charges are scaled to full, and GB solvent dielectric is scaled
to full.
\item [{States~7-8}] orientational and conformational restraints are removed.
\item [{State~8}] Temperature replica exchange molecular dynamics (TREMD)
or standard MD simulations of the unbound receptor and ligand with
flat-bottom harmonic restraints.
\end{description}
Instead of using \code{ISDDM.py} for sampling states 1 and 8, the
user may supply their own trajectories, generated with a sampling
method of their choice; otherwise, state 8 is simulated first. In
either case, the last frame from from state 8 is used as the initial
coordinates for all other simulations and is split into the unbound
ligand and host systems using \code{strip} command from \code{cpptraj}\cite{roe_ptraj_2013}
for states 1 to 4. Orientational and conformational restraints are
also constructed from the last frame of the state 8 trajectory. All
steps within in the workflow are handled using the Toil package \cite{vivian_toil_2017},
which is a pure Python workflow engine that allows the user run the
simulations in parallel.

\subsubsection{Conformational and Orientational Restraints\label{subsec:Conformational-and-Orientational}}

To maintain their respective conformations, \code{ISDDM.py} creates
separate sets of distance restraints for receptor and ligand. In each
case, harmonic distance restraints are assigned for every atom-pair
separated by less than 6 Å for the receptor and ligand. The reference
distances are taken from the last frame of the state 8 trajectory.
These restraints are applied in states 2 to 7.

\begin{figure}
\includegraphics[width=0.6\textwidth]{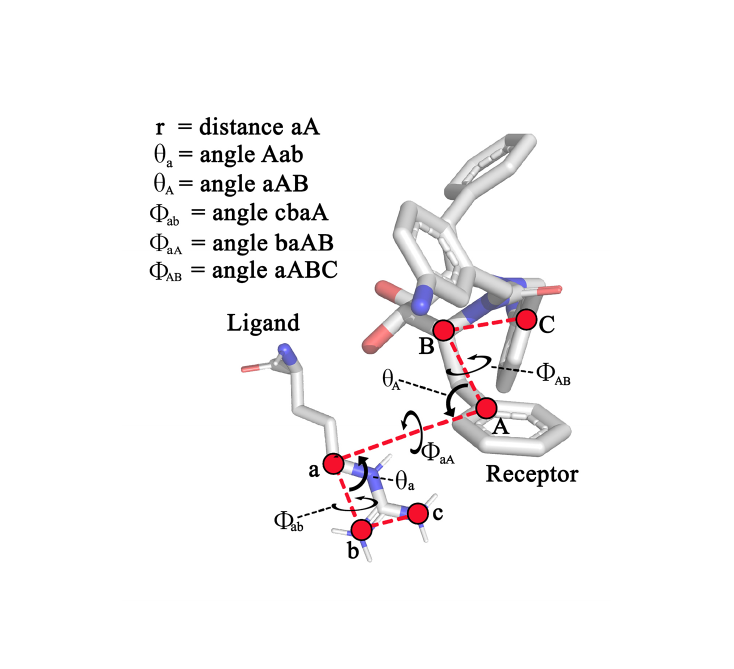}

\caption{Implementation of Boresch restraints for absolute binding free energy
calculations for an example receptor-ligand pair. Atoms $a$, $b$
and $c$ belong to the ligand (on the top right), while atoms $A$,
$B$ and $C$ belong to the protein (on bottom left).\label{fig:Boresch-restraints}}
\end{figure}

The orientation and distance between the receptor and ligand are restrained
using Boresch restraints \cite{boresch2003absolute}, which requires
a distance restraint, $r_{\text{aA}}$ , two angles, $\theta_{\text{A}}$
and $\theta_{\text{a}}$, and three torsions, $\phi_{\text{AB}}$,
$\phi_{\text{aA}}$ and $\phi_{\text{ba}}$ (\figref{Boresch-restraints}).
In principle, any three heavy atoms from the ligand ($\text{A}$,
$\text{B}$ and $\text{C}$) and receptor ($\text{a}$, $\text{b}$
and $\text{c}$) and be used, but the selection can impact the convergence
of the simulations. For example, if $\theta_{\text{A}}$ or $\theta_{\text{a}}$
are close to $180^{\circ}$, then the torsion angles $\phi_{\text{AB}}$,
$\phi_{\text{aA}}$ and $\phi_{\text{ba}}$ may experience large fluctuations.
Similarly, a large value of $r_{\text{aA}}$ may result in large fluctuations
of the ligand position in the binding pocket. While users can pre-select
all three atoms from ligand and receptor, \code{ISDDM.py} has an
automated procedure to mitigate any such problems.

\code{ISDDM.py} first selects atoms on the receptor and ligand for
the distance restraints $\text{A}$ and $\text{B}$. In \code{ISDDM.py}
the user can select three options for constructing the distances restraints
with the \code{restraint\_type}keyword.
\begin{enumerate}
\item For both the receptor and ligand, select the heavy atom closest to
the respective center of mass (COM).
\item Select atom A as the heavy atom nearest to the center of mass of the
ligand and receptor heavy atom, a, that is closest to A.
\item Lastly, the third option identifies the shortest distance between
any heavy atoms in the receptor and the ligand.
\end{enumerate}
Once the two distance atoms have been selected, \code{ISDDM.py} uses
an iterative approach to first identify heavy atoms b and B, then
c and C, such that angles $\angle Aab$, $\angle abc$, $\angle aAB$,
and $\angle ABC$ are between $80^{\circ}$ and $100^{\circ}$. If
\code{ISDDM.py} is unable to find suitable heavy atoms within the
desired range, it reduces the minimum angle and increases the maximum
angle by $1^{\circ}$ each. The range is expanded until suitable atoms
are found or the minimum angle reaches $10^{\circ}$, in which case
the program exits with an error. 

\subsubsection{Free Energy Analysis Using MBAR\label{subsec:Free-Energy-Analysis}}

\code{ISDDM.py} uses PyMBAR version 4.0\cite{shirts2008statistically}\cite{chodera_simple_2016}
to calculate free energy differences along the states 1-4 and 6-8
using the multistate Bennet acceptance ratio (MBAR)\cite{shirts2008statistically,bennettEfficientEstimationFree1976a}.
MBAR requires the potential energies of each coordinate trajectory
to be calculated with the potential function of all other states of
the same molecule. For example, potential energies for the coordinate
trajectory from state 2 need to be calculated for all other states
between 1 and 4. As each MD simulation completes, \code{ISDDM.py}
then calculates the potential energies of the resulting trajectory
in all other states using \code{sander} with \code{imin=5}.

The \code{mbar} module also provides error estimates for the calculated
binding free energy on the assumption that all samples are independent.
To ensure that the samples are not time correlated, which would underestimate
the error \cite{klimovich_guidelines_2015}, \code{ISDDM.py} uses
\code{timeseries} module of PyMBAR to remove correlated frames and
the relaxation period of the energy trajectories \cite{chodera_simple_2016}.

\subsection{Simulation details}

\subsubsection{System preparation}

Structures of cucurbit{[}7{]}uril (CB7), octa-acid (OAH), $\alpha$-cyclodextrin
(ACD), $\beta$-cyclodextrin (BCD), and Pillar{[}6{]}arene (WP6) host
molecules and their respective guests were obtained from the Taproom
GitHub repository \cite{slochowerTaproom2022} and host-guest complexes
created using Autodock Vina \cite{forli_computational_2016} (Figure
S1, Figure S2 and Figure S3). We prepared host molecules for docking
by using the \code{antechamber} and \code{tleap} programs \cite{case2023ambertools}
to assign AM1-BCC partial atomic charges \cite{jakalian_fast_2002}
and generalized Amber Force Field v2 (GAFFv2) force field parameters
\cite{wang2004development,cornell_second_1995}. We then relaxed the
host molecule structures with 5,000 steps of steepest descent energy
minimization with the \code{sander} program. For each host-guest
pair, Autodock Vina generated several poses of the fully flexible
ligand with a rigid, pre-relaxed host. In each case, the affinity
maps were 10 Å cubes, centered on the host center of mass. The lowest
energy ligand pose was selected for each host-guest complex for the
initial end-state configuration.

The host-guest complexes generated from docking were then parameterized
for use in AmberTools21, using AM1-BCC charges with GAFFv2. We applied
hydrogen mass repartitioning, in which mass of hydrogen atoms are
increased by a factory of 4 and the mass of the bonded heavy atoms
is decreased by an equivalent amount, to all complexes \cite{hopkins_long-time-step_2015}.

\subsubsection{\protect\code{ISDDM.py} parameters}

All production simulations were run with \code{ISDDM.py}, using \code{pmemd.mpi}
from Amber 21 \cite{AMBER21} for the molecular dynamics calculations.
Keywords specified in the YAML file are italicized, while MDIN keywords
are not.

\paragraph{Solvation models}

We set the desired GB model in MDIN files provided to \code{ISDDM.py}.
All five GB model variants available in AMBER 21 were used: Hawkins-Cramer-Truhlar
(HCT, \code{igb=1}) \cite{hawkins_pairwise_1995}; Onufriev-Bashford-Case
(OBC, \code{igb=2})\cite{onufriev_modification_2000}, OBC2 \cite{onufriev_exploring_2004}
(\code{igb=5}); GB-Neck (GBn, \code{igb=7}) \cite{mongan_generalized_2007};
and GBn2 (\code{igb=8}) \cite{nguyen2013improved}. The recommended
Born atomic radii sets mbondi, mbondi2, bondi and mbondi3 were used
for HCT, OBC/OBC2, GBn and GBn2 respectively and assigned in \code{tleap}
when the complexes were parameterized. The ionic concentration was
set to 0.3 M (\code{saltcon=0.3}) for all host guest systems except
for OAH, which used 0.15 M to better match experiment. The solvent-accessible-area
model of the non-polar solvation free energy was not used (\code{gbsa=0}).

\paragraph{Setup simulations for \protect\code{ISDDM.py} \label{subsec:End-state-Simulations}}

\code{ISDDM.py} carried out end-state simulations (states 1 and 8),
using replica exchange molecular dynamics simulation (REMD) \cite{sugita1999replicaexchange}.
As part of this procedure, \code{ISDDM.py} used \code{sander} or
\code{sander.MPI} to relax the initial structures with 5000 steps
of steepest descent energy minimization. The subsequent REMD calculations
used \code{sander.MPI}\textbf{ }with eight temperature copies\emph{,
\code{\emph{temperatures={[}300.00,}} \code{\emph{327.32,}} \code{\emph{356.62,}}
\code{\emph{388.05,}} \code{\emph{421.77,}} \code{\emph{457.91,}}
\code{\emph{496.70,}} \code{\emph{500.00{]}}},} for 2,500,000 exchange
attempts (\code{numexchg=2500000}) with exchanges attempted every
step (\code{nstlim=1}). The temperature distribution included in
the YAML file was determined by the \emph{Temperature generator for
REMD-simulations }server \cite{patriksson2008atemperature,spoel2023dspoelremdtemperaturegenerator}
for a target exchange probability of 0.35. The temperature was regulated
via Langevin dynamics (\code{ntt=3, gamma\_ln=1}). For all host-guest
bound end-state simulations, a flat-bottom distance restraint potential
was used to define the bound state. The harmonic restraint use a force
constance of $\unit[1]{\unitfrac{kcal}{mol\cdot\mathring{A}^{2}}}$
for separations between 5 Å and 10 Å. Beyond this, the potential was
linear. After the simulations completed, coordinate trajectories at
the target temperature (\emph{\code{\emph{target\_temperature=300}}})
were extracted by \code{pytraj}\cite{roe_ptraj_2013}. A total of
10,000 frames were saved for each of the end-states (\code{ntwx=250}).

Intermediate-state simulations (states 2 to 7) used the last frame
of REMD simulations in state 8 for the initial structure, which was
also the reference for the orientational and conformational restraints.
In state 7, the GB dielectric constant was reintroduced in sub-state
simulations with values of \code{gb\_extdiel\_windows=3.925, 7.85, 15.7, 39.25}
and \code{78.5}. Similarly, the ligand charges were reintroduced
with a single intermediate state with charges scaled by 0.5, \emph{\code{\emph{charges\_lambda\_windows={[}0.5, 1{]}}}}.
In all simulations, a time step of 4 fs (\code{dt=0.004}) was used
with SHAKE on all hydrogens (\code{ntc=2}), and the electrostatics
and GB cutoffs were set to $\unit[999]{\mathring{A}}$, (\code{cut=999, rgbmax=999}).
Except where noted, distance restraint force constants for both conformational
restraints and receptor-ligand distance restraints (\emph{\code{\emph{exponent\_\-conformational\_\-restraint}}})
ranged from $\unit[2^{-8}]{kcal/mol/\mathring{A}^{2}}$ to $\unit[2^{4}]{kcal/mol/\mathring{A}^{2}}$,
while orientational restraint force constants (\code{exponent\_\emph{\-}orientational\_\emph{\-}restraint})
ranged from $\unit[2^{-4}]{kcal/mol/rad^{2}}$ to $\unit[2^{8}]{kcal/mol/rad^{2}}$,
in 33 lambda-windows, spaced evenly on a $\log_{2}$-scale.

Five independent simulations were performed for each host-guest pair.
Each of these simulations use the same initial complex coordinates
but different random seeds were used, generating different trajectories
due to the initial velocities, Langevin dynamics and TREMD.

\paragraph{Sampling with TREMD}

To test the reliability and reproducibility of our results, we simulated
five independent copies of each host-guest system using the OBC model.
All five copies used different seeds for the pseudo-random number
generator, providing unique sequences of pseudo-random numbers for
velocities and replica exchange attempts. However, all copies used
the same force field, solvent parameters and initial starting coordinates.

A key metric for TREMD sampling is the number of round trips (RTs),
which is when a single replica traverses from the lowest temperature
to the highest and back again.\cite{roe2014evaluation} The reported
values reflect the average number of round trips per replica for the
complex (i.e., host and guest together), which exceeded 500, indicating
that the replicas visited the target temperature many times (Table
\ref{tab:Average-TREMD-statistics}). As is common, the variance in
number of exchanges required for a RTs was large, taking as little
as 22 exchanges or as much as 42,609. As we use the same temperatures
for all systems, we see that the mean RTs are correlated with the
number of atoms in the system. E.g., CB7 is the smallest host and
has the largest number of mean RTs.

\begin{table*}
\begin{tabular}{>{\centering}p{0.5in}>{\centering}p{0.5in}>{\centering}p{0.75in}>{\centering}p{0.75in}>{\centering}p{1in}>{\centering}p{1in}}
\toprule 
Systems & Mean RTs & Mean EX per RT  & $\sigma_{\text{EX}}$ per RT & Minimum EX for RT & Maximum EX for RT\tabularnewline
\midrule
BCD & 588 & 4291 & 4856 & 52 & 42609\tabularnewline
CB7 & 1377 & 1826 & 1955 & 22 & 15920\tabularnewline
OAH & 562 & 4463 & 3934 & 75 & 26606\tabularnewline
WP6 & 662 & 3830 & 3774 & 54 & 29051\tabularnewline
\bottomrule
\end{tabular}

\caption{TREMD round-trip (RT) and exchange (EX) statistics for each host.
Mean and standard deviation, $\sigma$, were averaged over all guest
molecules. \label{tab:Average-TREMD-statistics}}

\end{table*}

\subsection{Statistical metrics of absolute binding}

To assess the ABFE predictions of GB in this automation workflow protocol,
we applied several metrics, several of which are unique to the SAMPL
challenges \cite{muddanaSAMPL4HostGuest2014b}. Standard metrics included
the best-fit linear regression slope and Pearson correlation coefficient,
$R^{2}$. We also calculated the root mean squared error (RMSE),
\[
\text{RMSE}=\sqrt{\stackrel[i=1]{n}{\sum}\frac{\left(\Delta G_{i}^{\text{calc}}-\Delta G_{i}^{\text{exp}}\right)}{n}},
\]
mean absolute error (MAE),
\[
\text{MAE}=\frac{\sum_{i=1}^{n}\left|\Delta G_{i}^{\text{calc}}-\Delta G_{i}^{\text{exp}}\right|}{n},
\]
and mean signed error (MSE),
\[
\text{MSE}=\stackrel[i=1]{n}{\sum}\frac{\left(\Delta G_{i}^{\text{calc}}-\Delta G_{i}^{\text{exp}}\right)}{n},
\]
where $\Delta G^{\text{exp}}$ and $\Delta G^{\text{calc}}$ are the
experimental and GB predicated binding affinity values for guest molecule
$i$, and $n$ is the number of measurements. 

\section{Results\label{sec:Results}}

\subsection{Cycle convergence and numerical error}

\subsubsection{Uncertainty Estimates}

\code{ISDDM.py} provides numerical uncertainty estimates for the
calculated ABFE via the \code{pyMBAR} library \cite{klimovich_guidelines_2015}.
To evaluate the reliability of the uncertainty estimates, we calculated
the standard deviation of the mean ABFE for each host-guest pair based
from five independent simulations employing the OBC model (\tabref{Average-standard-errors}).
Overall, we see that the standard deviations of the mean ABFE across
the replicas for each host are below $\unit[1]{kcal/mol}$ except
for CB7, which is well within chemical accuracy. However, this is
significantly larger than the average uncertainty reported by \code{pyMBAR},
which may be due to time correlations in the TREMD data. Although,
\code{ISDDM.py} uses the \code{timeseries} module of \code{pyMBAR}
to remove time correlations from the intermediate simulation data,
this is not feasible for the TREMD, where time correlations may be
obscured as the replica simulations all contribute to sampling at
the target temperature in a random fashion. These results suggest
that the \code{pyMBAR} uncertainties are smaller than the variation
that can be expected and may not be a useful guide. Regardless \code{ISDDM.py}
can produce reliable results with uncertainties less than $\unit[1]{kcal/mol}$
with the settings used here. 

\begin{table}
\begin{tabular}{>{\centering}p{0.5in}>{\centering}p{1in}>{\centering}p{1in}}
\toprule 
System & Standard Deviation of Mean ABFE  & Mean MBAR Uncertainty Estimate\tabularnewline
\midrule
CB7 & 2.19 & 0.08\tabularnewline
WP6 & 0.64 & 0.33\tabularnewline
ACD & 0.46 & 0.11\tabularnewline
BCD & 0.43 & 0.003\tabularnewline
OAH & 0.41 & 0.14\tabularnewline
\bottomrule
\end{tabular}

\caption{Standard deviations of ABFE and mean MBAR error estimates from independent
copies of simulations with OBC (\protect\code{igb=2}) in kcal/mol.
Standard deviations of the mean ABFE were computed for each host-guest
pair and then averaged over all pairs for each host. \protect\code{pyMBAR}
uncertainties were computed for each replica and averaged for each
host. \label{tab:Average-standard-errors}}
\end{table}

\subsubsection{Intermediate state overlap}

When using any perturbation approach, such as FEP or MBAR, a poor
phase-space overlap between states can result in an inaccurate free
energy estimate. Overlap matrices are an effective method to assess
phase-space overlap \cite{klimovich_guidelines_2015}. The matrix
shows the probability that a sample from state $i$ could be observed
in state $j$, indicating the degree of overlap between both states.
To obtain reliable free energy estimates, the overlap matrix should
at least be tridiagonal, and all $\left(i,i+1\right)$ and $\left(i-1,i\right)$
elements should be greater than 0.03 \cite{klimovich_guidelines_2015}. 

\begin{figure}
\includegraphics[width=0.5\textwidth]{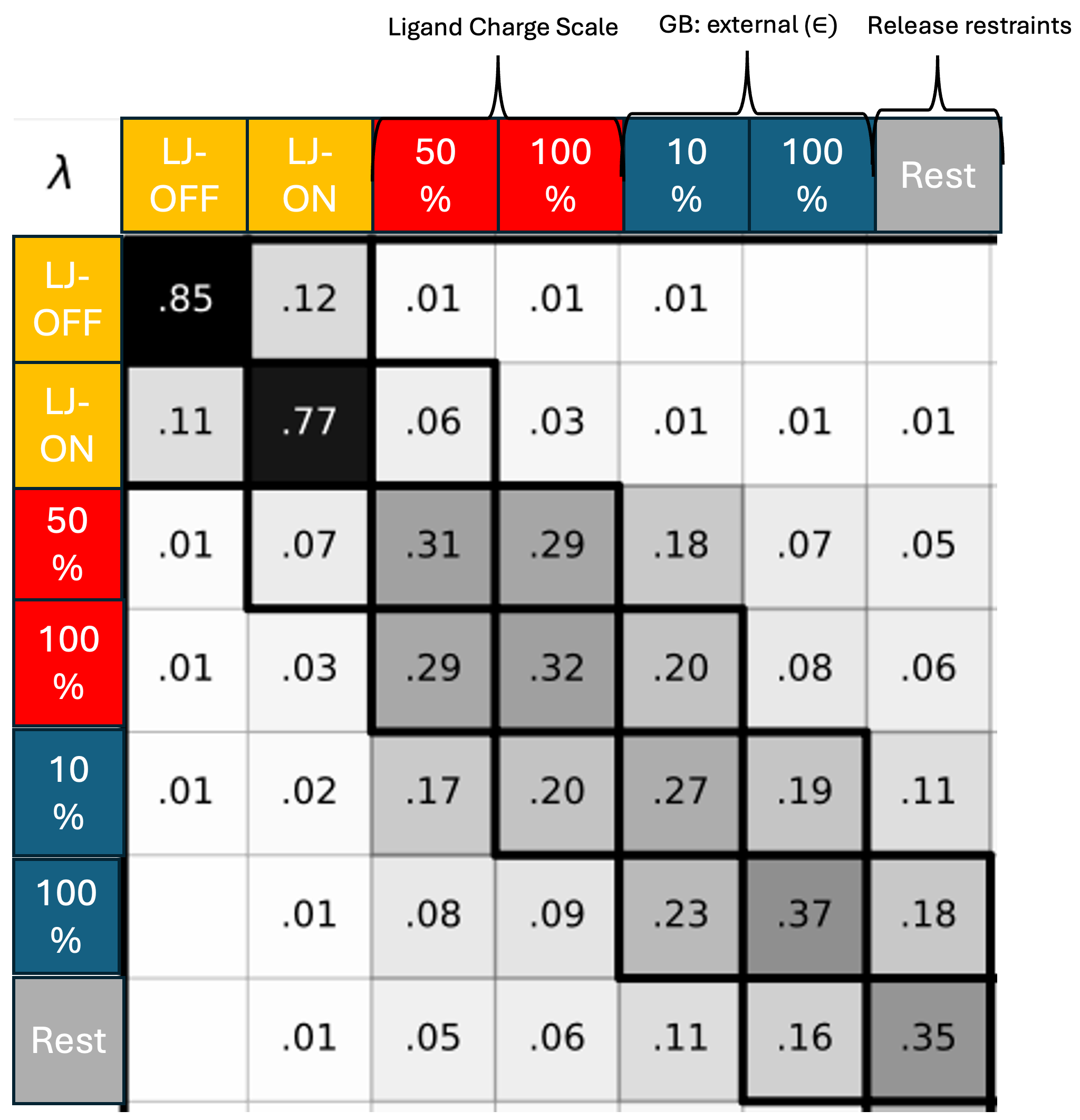}

\caption{Partial overlap matrix calculated for the cb7 complexed with adn using
the OBC model, from state 6 with no host-guest interactions to state
7, where the host and guest are fully interacting.\label{fig:Overlap-matrix.}
Entire matrix can be found in Figure S4. }
\end{figure}

\Figref{Overlap-matrix.} shows the overlap matrix for states 6 to
7 (\secref{Implementation-Details}) for the CB7-adn complex. The
(No-LJ,No-LJ) cell shows that if sampling is done in the No-LJ state
and analyzed in same state, there is a $0.85$ overlap. I.e., 85\%
of equilibrated configurations are assigned to the No-LJ state and
not adjacent states. The probability of finding a microstate sampled
from the No-LJ state in the LJ-On state is 0.12, and so is finding
the microstate sampled from state LJ-On in state No-LJ. Significantly,
the high overlap means that conformational restraints are an effective
replacement for soft-core LJ potentials. 

In the next thermodynamic step, where the ligand charge is turned
back on to 50\% of the net charge, sampling in the LJ-On state and
analyzing in the 50\% ligand charge state reveals a 6\% overlap. The
subsequent intermediate windows, where the ligand becomes fully charged
and the scaling of the GB external dielectric is applied, show good
overlaps, indicating that fewer windows may be sufficient. Specifically,
in the case of a 50\% ligand charge, there is an 18\% overlap with
the GB external dielectric of 10\%, suggesting that the fully charged
ligand state could potentially be skipped. 

Overlaps for adjacent states were generally well over 0.03 (Figure
S4). In fact, many windows, particularly for the weak restraint constants,
could be omitted, thereby further reducing the computational requirements.

\subsubsection{Effect of restraint magnitude}

Because the change in free energy is path independent, conformational
and orientational restraints should have no impact on the value of
ABFE calculated, as long as they are introduced and removed as part
of the thermodynamic cycle. To ensure that this is the case, there
must be sufficient phase-space overlap between the unrestrained end-states
and the adjacent intermediate states with the weakest conformational
and orientational restraints. For example, when removing conformational
restraints between states 7 and 8, the intermediate state with the
minimum restraint force constant must be low enough to provide sufficient
overlap with the end-states, which have no such restraints. To determine
the lowest restraint constants required, we reduced the force constants
until sufficient overlap was found with the end-states. We found that
a distance restraint force of $\unit[2^{-8}]{kcal/mol/\mathring{A}^{2}}$
with an orientational restraint force of $\unit[2^{-4}]{kcal/mol/rad^{2}}$
typically gave an overlap of about 0.18.

Conversely, the maximum magnitude of the force constants has a strong
effect on the number and type of intermediate states required in the
cycle. For example, if the force constants for restraints are too
small, steric overlaps between the ligand and receptor may occur when
LJ interactions are turned off, which would require the introduction
of soft-core LJ intermediate states. Alternatively, if force constants
for restraints are large enough, steric overlaps can be prevented
and the need for soft-core LJ intermediate states eliminated. However,
there is a trade-off to using larger force constants, as they require
more intermediate states to introduce and remove the restraints.

To assess the impact of the maximum force constants for conformational
and Boresch restraints, we calculated the ABFE of CB7 with all guests
using two sets of maximum force constants. For the weakest restraints,
11 windows were used, evenly spaced on a $\log_{2}$-scale from $\unit[2^{-8}]{kcal/mol/\mathring{A}^{2}}$
to $\unit[2^{2}]{\unitfrac{kcal}{mol\cdot\mathring{A}^{2}}}$ for
distance restraints and $\unit[2^{-4}]{kcal/mol/rad^{2}}$ to $\unit[2^{6}]{kcal/mol/rad^{2}}$
for orientational restraints. For the stiffest restraints, four additional
windows with stiffer restraints were added evenly spaced on a $\log_{2}$-scale
to give maximum values of $\unit[2^{4}]{\unitfrac{kcal}{mol\cdot\mathring{A}^{2}}}$
and $\unit[2^{8}]{kcal/mol/rad^{2}}$. In both cases, host-guest LJ
interactions were turned on in a single step, but additional electrostatic
windows were required for the weak-restraint calculation to ensure
sufficient overlap. To test this, five independent replica ABFE calculations
were carried out on all CB7-guest pairs with both sets of restraints
and no systematic difference was observed (\tabref{GB-RISM_BFEs}),
demonstrating that our calculations are path independent and insensitive
to the restraint magnitudes.

\begin{table}
\begin{centering}
\begin{tabular}{>{\centering}p{0.4in}>{\centering}p{0.7in}>{\centering}p{0.7in}>{\centering}p{0.7in}}
\toprule 
Guest & Stiff restraints & Weak restraints & Difference\tabularnewline
\midrule
C1 & -14.4(1) & -14.1(1) & -0.3\tabularnewline
C2 & -13.9(2) & -13.8(3) & -0.1\tabularnewline
C3 & -11.4(1) & -11.1(3) & -0.3\tabularnewline
C4 & -13.9(1) & -13.7(1) & -0.2\tabularnewline
C5 & -9.8(0) & -10.0(1) & 0.2\tabularnewline
C6 & -15.3(1) & -15.2(1) & -0.1\tabularnewline
C7 & -16.6(2) & -16.4(1) & -0.2\tabularnewline
C8 & -18.1(1) & -17.9(0) & -0.2\tabularnewline
C9 & -19.5(1) & -19.8(1) & 0.3\tabularnewline
C10 & -15.6(1) & -15.6(1) & 0.0\tabularnewline
C11a & -18.8(1) & -18.6(0) & -0.2\tabularnewline
C11b & -19.0(1) & -18.7(1) & -0.3\tabularnewline
C12 & -20.0(1) & -20.0(1) & 0.0\tabularnewline
C13 & -23.4(1) & -23.2(1) & -0.2\tabularnewline
C14 & -21.7(1) & -21.6(2) & -0.1\tabularnewline
Average & -15.3 & -15.2 & \tabularnewline
\bottomrule
\end{tabular}
\par\end{centering}
\caption{The effect of stiff and weak restraints on calculated binding free
energy. Stiff restraints uses a maximum force constant of $\unit[2^{4}]{\unitfrac{kcal}{mol\cdot\mathring{A}^{2}}}$
for distance and $\unit[2^{8}]{\unitfrac{kcal}{mol\cdot rad^{2}}}$
for angle and torsion restraints. Weak restraints uses a force constant
of $\unit[2^{2}]{\unitfrac{kcal}{mol\cdot\mathring{A}^{2}}}$ for
distance and $\unit[2^{6}]{\unitfrac{kcal}{mol\cdot rad^{2}}}$ for
angle and torsion restraints. Units are in kcal/mol and standard error
in the last digit is given in parentheses, calculated from five independent
replicas. \label{tab:GB-RISM_BFEs}}
\end{table}

\subsection{Parallel scaling}

\begin{figure}
\includegraphics{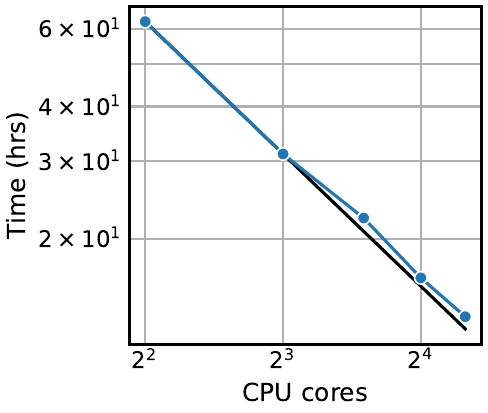}

\caption{Parallel scaling of intermediate steps of \protect\code{ISDDM.py}
for the 150-atom CB7-ADN complex. The black line shows linear scaling
relative to four CPU-cores and the blue line is the actual measured
time.\label{fig:-processor-scaling}}
\end{figure}

As \code{ISDDM.py} utilizes the \code{Toil} workflow management
system \cite{vivian_toil_2017}, users are able to employ multiple
processors to speedup ABFE calculations. As all intermediate states
(2-7) are independent, all MD simulations, post-analysis calculations
and MBAR computations for these states can be parallelized over the
available processors identified by the user. \Figref{-processor-scaling}
shows the processor time to complete all intermediate states (states
2-7) for the 150-atom CB7-ADN complex system. When intermediate states
are predefined by the user, the automation application achieves near-perfect
scaling as the number of processes increases. In practice, the user
should ensure that lambda windows are evenly spaced on a $\log_{2}$-scale
to maintain linear scaling. End-state calculations are also parallelized,
and \code{ISDDM.py} will automatically run the end-state calculations
concurrently with each other on the available CPU-cores. However,
conformational and Boresch restraints are taken from the bound-state
simulation, which must complete before intermediate states can run.

\subsection{Absolute binding free energy with different GB models}

\begin{figure*}
\includegraphics{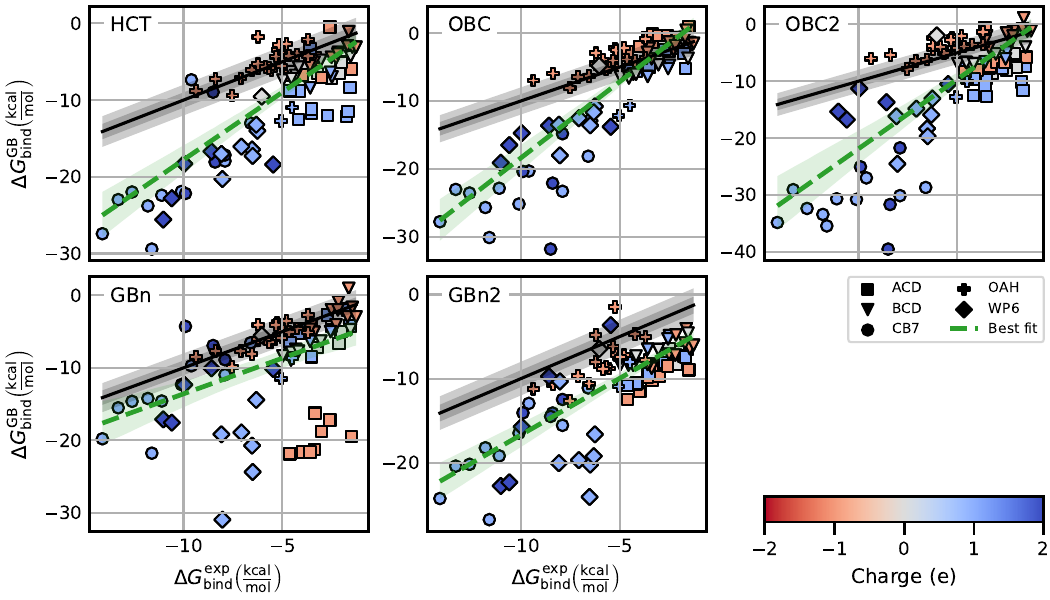}

\caption{Comparison of calculated and experimental absolute binding free energies
for all systems with five different GB models: HCT, OBC, OBC2, GBn,
and GBn2. The black solid line is the experimental identity. The dashed
green line indicates the linear regression fitting for host guest
systems.\label{fig:effect_gb_models}}
\end{figure*}

As the the solvent environment is a critical determinant molecular
binding affinity, we first evaluate the predictions of all tested
generalized Born models across the five host-guest systems (\tabref{All-GB-uncorrected}
and \figref{effect_gb_models}).Among the tested GB models, OBC consistently
delivered the best overall performance for all host-guest systems.It
both had the best metrics when all systems were considered together,
except for the slope of the best fit line, and provided the best or
second best metrics for each host system, except OAH. However, systematic
errors that correlate with ligand net charge severely affected the
accuracy.

Both HCT and OBC2 show qualitatively similar binding free energy distributions
to OBC (\figref{effect_gb_models}), though both show larger error
for positively charged ligands Specifically, in all three models,
the ABFE of positively charged ammonium groups tend to be underestimated,
while negatively charged carboxyl groups are overestimated in ABFEs
energies. Comparing all systems to experimental data HCT and OBC2
yielded root mean square errors (RMSE) of $\unit[6.42]{kcal/mol}$
and $\unit[6.12]{kcal/mol}$respectively. OBC also achieved the highest
Pearson correlation coefficient with experiment $R^{2}\approx0.86$,
indicating strong agreement with observed trends. 

GBn performed well, except for two sets of outliers with significantly
underestimate ABFEs (\figref{effect_gb_models}): ligands with $-1e$
net charge bound to WP6 and ligands with $+1e$ net charge bound to
ACD. The WP6 outlier guest molecules contained hydrocarbon chains
with ammonium functional groups, while the ACD outlier guest molecules
were carboxylate functional groups. No structural similarity was identified
between the two sets of guest molecules. However further analysis
showed strong and constant hydrogen bonding between the negatively
charged carboxylate groups of the guest ligands with the hydroxyl
groups of host ACD. Particularly, three stable hydrogen bonds were
seen all through the simulation between the alcohol groups of ACD
and the oxygens of carboxylate from guests. This is likely what makes
the GBn model overestimate binding affinities for these ligands as
noted by the systematic deviation from the experimental identity line
shown in \figref{effect_gb_models}. This type of behavior seems to
apply only to ACD which has smaller ring size six glucose units compared
to BCD seven glucose units although they are chemically similar. The
GBn model handled negatively charged guests more accurately with BCD,
as evidenced by predictions within 2.4 kcal/mol of experiment. This
suggests a potential system-dependent limitation of the GBn model,
where subtle host structural differences (e.g., cavity size, hydrogen
bond geometry) may amplify or dampen electrostatic over-stabilization.
Lastly, the GBn2 (\figref{effect_gb_models}) tends to underestimate
ABFE for all ligands. While positively charged ligands are underestimated
more, the effect is much smaller than in the other four models.

\begin{table*}
\begin{tabular}{ccccccc}
\toprule 
System & GB model & RMSE$_{\text{std}}$ & MAE & $R^{2}$ & Slope (m) & $\tau$\tabularnewline
\midrule
All-Systems & HCT & 6.42 & 4.8 & 0.79 & 0.35 & 0.48\tabularnewline
 & OBC & 6.12 & 3.96 & 0.86 & 0.33 & 0.65\tabularnewline
 & OBC2 & 9.32 & 6.34 & 0.76 & 0.23 & 0.43\tabularnewline
 & GBn & 6.76 & 4.14 & 0.47 & 0.22 & 0.42\tabularnewline
 & GBn2 & 6.18 & 5.26 & 0.77 & 0.45 & 0.53\tabularnewline
ACD & HCT & 5.59 & 4.71 & 0.0 & 0.0 & 0.02\tabularnewline
 & OBC & 1.79 & 1.45 & 0.4 & 0.2 & 0.30\tabularnewline
 & OBC2 & 5.78 & 5.43 & 0.14 & 0.05 & 0.23\tabularnewline
 & GBn & 10.2 & 7.22 & 0.28 & 0.04 & 0.30\tabularnewline
 & GBn2 & 6.08 & 6.01 & 0.85 & 0.52 & 0.69\tabularnewline
BCD & HCT & 2.43 & 2.20 & 0.77 & 0.50 & 0.69\tabularnewline
 & OBC & 1.62 & 1.41 & 0.80 & 0.40 & 0.62\tabularnewline
 & OBC2 & 1.95 & 1.68 & 0.79 & 0.35 & 0.62\tabularnewline
 & GBn & 2.43 & 2.13 & 0.84 & 0.36 & 0.68\tabularnewline
 & GBn2 & 3.96 & 3.85 & 0.75 & 0.59 & 0.55\tabularnewline
CB7 & HCT & 10.57 & 9.69 & 0.70 & 0.25 & 0.60\tabularnewline
 & OBC & 13.84 & 13.28 & 0.44 & 0.22 & 0.40\tabularnewline
 & OBC2 & 20.91 & 20.49 & 0.32 & 0.16 & 0.31\tabularnewline
 & GBn & 3.89 & 2.88 & 0.78 & 0.35 & 0.62\tabularnewline
 & GBn2 & 7.49 & 6.92 & 0.83 & 0.41 & 0.71\tabularnewline
OAH & HCT & 2.51 & 1.56 & 0.38 & 0.19 & 0.42\tabularnewline
 & OBC & 2.61 & 1.90 & 0.31 & 0.17 & 0.40\tabularnewline
 & OBC2 & 2.77 & 2.10 & 0.24 & 0.13 & 0.34\tabularnewline
 & GBn & 1.91 & 1.28 & 0.51 & 0.33 & 0.40\tabularnewline
 & GBn2 & 3.00 & 2.71 & 0.60 & 0.33 & 0.45\tabularnewline
WP6 & HCT & 10.03 & 9.63 & 0.76 & 0.34 & 0.51\tabularnewline
 & OBC & 6.32 & 5.98 & 0.68 & 0.35 & 0.59\tabularnewline
 & OBC2 & 8.65 & 7.67 & 0.21 & 0.07 & 0.21\tabularnewline
 & GBn & 11.14 & 9.01 & 0.10 & 0.03 & 0.06\tabularnewline
 & GBn2 & 10.35 & 8.78 & 0.38 & 0.11 & 0.38\tabularnewline
\bottomrule
\end{tabular}

\caption{Comparison of all host guest systems with different IGB models (\protect\code{igb=1},
\protect\code{igb=5},\protect\code{igb=7},\protect\code{igb=8}).
Units for RMSE\_std and MAE are in kcal/mol. The best result for each
system is highlighted with bold text. \label{tab:All-GB-uncorrected}}
\end{table*}

\subsection{Host-guest absolute binding free energies with OBC}

\begin{figure*}
\includegraphics{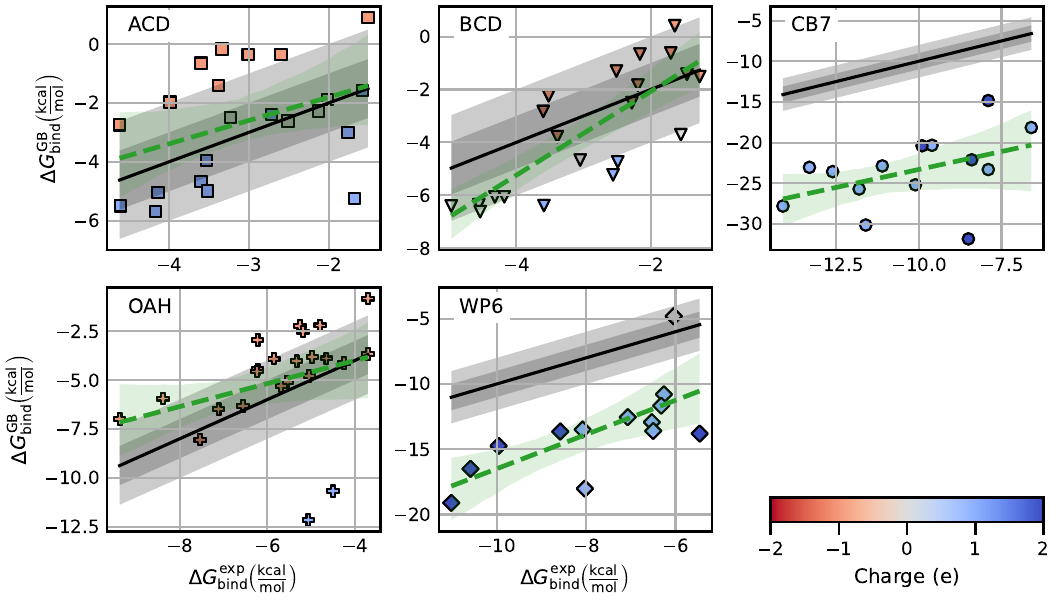}

\caption{Comparison of calculated and experimental absolute binding free energies
for all systems with the OBC solvation model. Coloring and symbols
are the same as \figref{effect_gb_models}.\label{fig:Calculated-absolute-binding}}
\end{figure*}

As the OBC solvation model has the best overall results among the
GB models, but there is a large variance in the metrics for the different
hosts, we examine each the behavior of each host in turn (\Figref{Calculated-absolute-binding}
and \Tabref{All-GB-uncorrected}). For example, individual systems
had RMSE\textsubscript{std} ranging from $\unit[1.62]{kcal/mol}$
(BCD) to $\unit[13.84]{kcal/mol}$ (CB7) and Pearson correlation coefficients
from 0.04 (ACD) to 0.80 (BCD). Much of this variance can be attributed
to presence of carboxyl or ammonium groups in the ligand, which results
in under estimation of the ABFE for all ligands with ammonium (positively
charged) and over estimation for most ligands with a carboxyl (negatively
charged), regardless of host (\Figref{Calculated-absolute-binding}).
However, the presence of carboxyl groups on the host has little impact
on the ABFE, despite the fact that CB7, ACD and BCD have neutral net
charge do not contain carboxyls while WP6 and OAH have net charges
of $-12e$ and $-8e$ due to carboxyls.

\subsubsection{$\alpha/\beta$-cyclodextrin}

ACD and BCD are chemically similar, differing in only the number of
repeated subunits, and have similar binding metrics, though they share
only 10 guests in their test sets. Both ACD and BCD had 22 guests
in their test sets, of which 10 were common between the two. In both
cases, guest net charges varying from $+1e$ to $-1e$ and included
common functional groups found in drug-like molecules such as ammonium,
alcohol and carboxyl. Consistent with the overall trend, both ACD
and BCD results show a clear bias in the error of the ABFE due to
ligand charges ( \figref{Calculated-absolute-binding}). This is most
dramatically seen with ACD, where the ABFE is over estimated for all
negatively charged ligand, underestimated for all positively charged
ligands, and near experiment for all neutral ligands. In all cases,
carboxylates are the source of negative net charges, while ammonium
groups are the source of positive net charges (Figure S2).

ACD and BCD have the lowest $\text{RMSE}$ and MAE values of all host
(\Tabref{All-GB-uncorrected}), likely due to the fact that the experimental
ABFEs are higher than for the other hosts and the error for all systems
is correlated with the magnitude of the experimental ABFE (\Figref{Calculated-absolute-binding}).
However, the slopes of the best fit lines (0.2 for ACD and 0.4 for
BCD) and the Pearson correlation coefficients ($R^{2}=0.4$ for ACD
and $R^{2}=0.80$ for BCD) were quite different. This is due to several
factors, including the slightly larger dynamic range of binding free
energies of BCD, than ACD, that ACD's negatively charged ligands have
a lower average ABFE, and that guest ligand (nmb) is a prominent outlier
for ACD (calculate: $\unit[-5.3]{kcal/mol}$, experimental: $\unit[-1.7]{kcal/mol}$).

\subsubsection{Cucurbit{[}7{]}uril}

We observed the correlation with experiment for CB7 ($R^{2}=0.44$),
which was paired with 14 cationic guests with charges from $+1e$
to $+2e$ (\figref{Calculated-absolute-binding}d and \Tabref{All-GB-uncorrected}).
The strong binding observed, appears to be due to favorable polar
interactions between the ammonium and carbonyl rich parts of CB7.

\subsubsection{Octa-acid}

OAH has a deep 10 Å hydrophobic pocket and eight water-solubilizing
carboxyl groups at the outer surface, which gave it a negative net
charge of $-8e$ (\figref{Calculated-absolute-binding}e). The OAH
host carboxylates remain well-solvated throughout the binding process,
preventing any systematic overbinding or desolvation penalty for negatively
charged guests. 23 of the guest molecules had a net charge of $-1e$
and two (tzh and hxa) had a net charge of $+1e$. This lead to poor
over all metrics (\Tabref{All-GB-uncorrected}), especially the Pearson
correlation coefficient ($R^{2}=0.31$), as the positively charged
ligands were large outliers from the rest of the data. However, these
positively charge ligands follow the same trend compared to experiment
as as all other positively charged ligands. If omitted, the correlation
improves to $R^{2}=0.75$.

\subsubsection{Pillar{[}6{]}ene}

The 13 guest ligands were tested against WP6, which had a net negative
charge of $-12e$, as shown in \figref{Calculated-absolute-binding}f.
Of the 13 guests, 12 had net charges of $+1e$ or $+2e$, due to ammoniums,
and one, G8, had a net neutral charge. Despite the difference in charge
of the hosts, we see the binding free energy for all positively charged
guest were underestimated to a similar magnitude as observed for CB7.
G8 is Zwitterionic, with a quaternary amide providing a localized
positive charge and a sulfate group contributing a negative charge,
neither of which have close interactions with carboxyl groups of WP6.
Omitting the G8 ligand from the data set yields binding metrics similar
to those for CB7, giving $R^{2}=0.74$ and $\text{RMSE}_{\text{std}}=6.57$.

\subsection{Linear correction for ammonium and carboxyl groups}

Despite the overall reasonable performance of the DDM/GB workflow,
binding free energies predicted by all five GB models exhibited a
systematic charge\nobreakdash-dependent bias. Ligands bearing ammonium
functionalities (net charge +1e) were consistently underestimated,
whereas those containing carboxylate groups (net charge --1e) were
systematically overestimated (Fig. 6, 7). This host\nobreakdash-independent
trend indicates source of these systematic errors lies in the parameterization
of these functional groups, and suggest a simple correction can mitigate
these errors without reparameterization:
\begin{equation}
\Delta G_{\text{bind}}^{\text{corr}}=(1+a)\Delta G_{\text{bind}}^{\text{GB}}+b,\label{eq:charge-correction-1}
\end{equation}
where $\Delta G_{\text{bind}}^{\text{GB}}$ is the raw GB\nobreakdash-predicted
binding free energy, and $a$ and $b$ are empirical coefficients
determined separately for ammonium and carboxylate ligands. Parameters
were obtained via five\nobreakdash-fold cross\nobreakdash-validation
over the combined host--guest dataset; fitted values are reported
in \tabref{Table-4.2:-Hydration}.

\begin{figure*}
\includegraphics{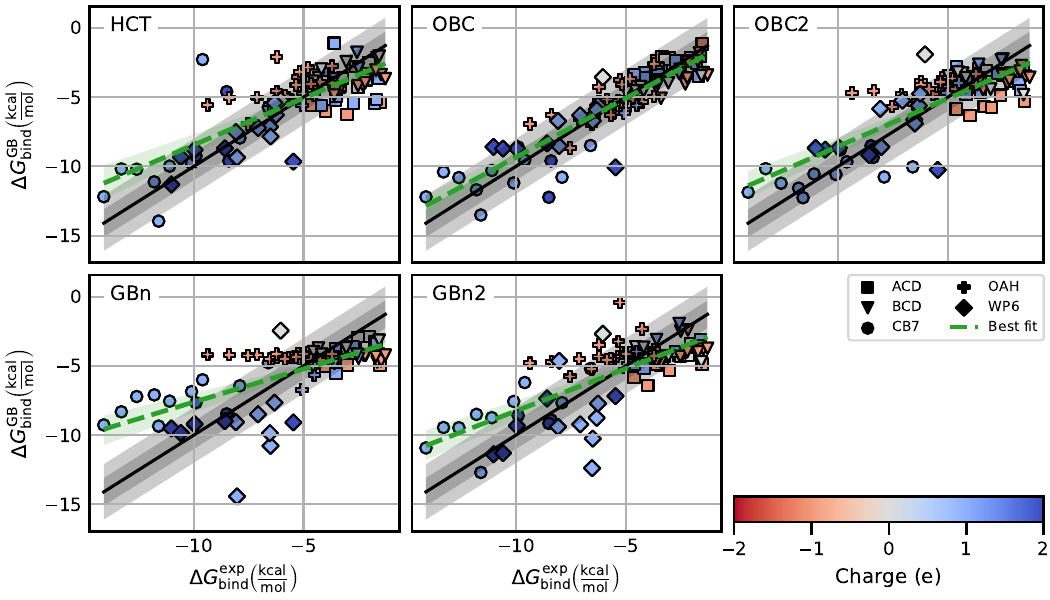}

\caption{Comparison of corrected calculated and experimental absolute binding
free energies for all systems with five different GB models: ) HCT,
OBC2, GBn, and GBn2. The black solid line is the experimental identity.
The dashed green line indicates the linear regression fitting for
host guest systems.\label{fig:correction_abfe}}
\end{figure*}

Application of this correction yields substantial improvements across
all GB models (\tabref{CHA_metrics}; \tabref{All-GB-uncorrected}).
For example, the OBC model (\code{igb=2}) RMSE decreases from 6.12
to 1.31 kcal/mol, and the regression slope increased from 0.33 to
0.92. HCT, OBC2, GBn, and GBn2 exhibit analogous enhancements, with
corrected RMSEs ranging from 1.86 to 2.25 kcal/mol and slopes approaching
unity (0.92--0.95) (\figref{correction_abfe}). However, this also
reduces the dynamic range of the calculated values at the cost of
fine-grained order of ligands with similar binding free energies.
As a result, correlation with experiment decreased from $R{{}^2}=0.47-0.86$
to $R{{}^2}=0.45-0.81$. These improvements suggest that force field
errors for specific functional groups can be mitigated with a simple
two-parameter correction.

Although the linear correction in the GBn model effectively addresses
the dominant source of error and improves binding metrics, errors
are increased for certain host--guest pairs due to outliers in the
original binding calculations (\figref{effect_gb_models}). The binding
mode of all negatively charged ligands displayed strong hydrogen\nobreakdash-bonding
interactions with the hydroxyl groups of ACD throughout the end\nobreakdash-state
simulation, which likely underlies the GBn--carboxylate outliers.
Similarly, in the GBn predictions, unusually strong binding was observed
for all guests with $\lyxmathsym{\textendash}1e$ charge in WP6, where
multiple ammonium--carboxyl interactions form between the guest ammonium
groups and WP6 carboxyl groups; this network is not possible in OAH,
owing to the larger separation between its carboxyl groups. In contrast,
GBn model results for guests with a $\lyxmathsym{\textendash}2e$
charge did not exhibit multiple ammonium--carboxyl interactions,
likely because the two negative charges reside on opposite ends of
these ligands, leading to intramolecular competition. These results
for GBn highlight that charge and solvent model errors can introduce
host\nobreakdash-specific hydrogen\nobreakdash-bonding and ion\nobreakdash-pair
geometry that cannot be accounted for by the simple linear correction.

\begin{table}
\begin{tabular}{cccc}
\toprule 
IGB model & Charge & a & b\tabularnewline
\midrule
HCT & -1 & -0.42 & -1.56\tabularnewline
 & 0 & -0.28 & -1.06\tabularnewline
 & 1 & -0.52 & 0.60\tabularnewline
 & 2 & -0.79 & -5.10\tabularnewline
OBC & -1 & -0.22 & -2.04\tabularnewline
 & 0 & -0.44 & -1.07\tabularnewline
 & 1 & -0.6 & -1.29\tabularnewline
 & 2 & -0.95 & -8.14\tabularnewline
OBC2 & -1 & -0.65 & -2.95\tabularnewline
 & 0 & -1.12 & -4.31\tabularnewline
 & 1 & -0.667 & -0.45\tabularnewline
 & 2 & -1.00 & -9.10\tabularnewline
GBn & -1 & -0.99 & -4.3\tabularnewline
 & 0 & -0.49 & -0.21\tabularnewline
 & 1 & -0.66 & -2.65\tabularnewline
 & 2 & -0.84 & -7.33\tabularnewline
GBn2 & -1 & -0.66 & -1.59\tabularnewline
 & 0 & -0.62 & -0.87\tabularnewline
 & 1 & -0.52 & 0.13\tabularnewline
 & 2 & -0.74 & -5.33\tabularnewline
\bottomrule
\end{tabular}

\caption{Hydration asymmetry linear corrections parameters, fit with five-fold\textbf{
}cross validation. Units of charge are $e$ and units of $b$ are
kcal/mol. \label{tab:Table-4.2:-Hydration}}
\end{table}

\begin{table}
\begin{tabular}{cccccc}
\toprule 
Model & RMSE & MAE & $R^{2}$ & Slope & $\tau$\tabularnewline
\midrule
HCT & 1.86 & 1.42 & 0.62 & 0.93 & 0.50\tabularnewline
OBC & 1.31 & 1.00 & 0.81 & 0.95 & 0.74\tabularnewline
OBC2 & 1.77 & 1.42 & 0.66 & 0.95 & 0.50\tabularnewline
GBn & 2.25 & 1.77 & 0.45 & 0.95 & 0.41\tabularnewline
GBn2 & 2.03 & 1.68 & 0.55 & 0.92 & 0.49\tabularnewline
\bottomrule
\end{tabular}

\caption{Comparison of all host guest systems after the application of equation
4.1. Units for RMSE and MAE are in kcal/mol. \label{tab:CHA_metrics}}
\end{table}

\section{Discussion\label{sec:Discussion}}

\subsection{Solvation models\label{subsec:Solvation-models}}

Many of the complexities of calculating ABFEs with explicit solvent
using DDM can be avoided or mitigated by adopting the GB implicit
solvent model. As implicit solvent models integrate out the solvent
degrees of freedom, steric overlaps between solute and solvent are
avoided, removing the need for soft-core LJ potentials. GB also enhances
computational speed and sampling efficiency \cite{anandakrishnan_speed_2015,onufriev2019generalized}.
Furthermore, implicit solvent models with open boundaries, such as
GB, naturally avoid periodic artifacts that arise from changes in
net charge or unit cell size. However, the results of GB-based ABFE
calculations depend critically on two factors: the cavity radii used
in the model and the atomic partial charges applied to the solutes.

In this study, we conducted a benchmark of all GB models available
in the AmberTools molecular modeling suite for comprehensive comparison.
Our calculations show that ABFEs from all GB solvent models we tested
correlate moderately well with experiment. Despite this, even the
best model (OBC) demonstrated an imbalance between direct electrostatic
interactions and solvent-mediated screening, resulting in systematic
inaccuracies for charged host--guest complexes. The GB cavity radii
fail to correctly balance direct Coulomb interactions with dielectric
effects, leading to systematic overestimation of binding affinities.
Over-binding is most pronounced in systems such as cucurbit{[}n{]}urils
with cationic guests because strong electrostatic attractions to electronegative
carbonyl portals are not sufficiently screened, leading to errors
on the order of tens of kcal/mol.

One possible remedy, explored by Setiadi et al \cite{setiadi2024tuningpotential}
for OBC2, is to reoptimize the GB cavity radii. They did this by refitting
five key atomic radii (hydrogen, nitrogen, carbon, oxygen, and hydrogen
bound to nitrogen) and realized a dramatic improvement in binding
free energy predictions, lowering the root-mean-square error (RMSE)
from about 21 kcal/mol to only 3 kcal/mol over a wide training set.This
result has shown that the inaccuracies of standard cavity radii are
major contributors to the electrostatic imbalances built into GB models.
However, this optimization for binding led to overestimation of hydration
free energies of the unbound ligands and receptor, suggesting limitations
of the GB framework.

An alternative approach is to examine the impact of the partial charge
model on ABFE accuracy. Using four CB7 ligands examined by Setiadi
et al \cite{setiadi2024tuningpotential}, we explored four different
combinations of partial charges models with GAFF2 and OBC2: AM1-BCC
(our results), restrained electrostatic potential (RESP) from PyRED
\cite{bayly_well-behaved_1993,vanquelef_red_2011}, and charges as
given in the Taproom repository. The AMI-BCC charge model was developed
to produce RESP-like charges at a lower computational cost \cite{jakalian_fast_2002}.
We note that we did not test the more recent ABCG2 charge method,
which is now recommend for the use with GAFF2\cite{he_fast_2020}.

We found that changing the charge assignment protocol significantly
altered the predicted \textgreek{\textDelta}G values (Table \ref{tab:Comparison-of-charge-models}).
For example, exchanging AM1-BCC for RESP charges lead to differences
in ABFE of up to $\unit[18]{kcal/mol}$. Similarly, our original results
for OBC2 were 18 to 47 kcal/mol higher than those obtained by Setiadi
et al\cite{setiadi2024tuningpotential} for their unoptimized calculations,
which used charges and radii generated in their custom workflow. When
we repeated calculations with structures and charges from the Taproom
GitHub repository, we found that the recalculated values were similar
to the highly negative \textgreek{\textDelta}G values reported by
Setiadi et al, confirming the sensitivity of GB calculations to charge
assignment. While it is possible the discrepancy between our results
and the unoptimized results of Setiadi et al is due to differing free
energy methods---they used the attach--pull--release (APR) approach
\cite{velezvega_overcoming_2013}---a more likely cause is the underlying
charge model as shown by these calculations

These results demonstrate that ABFE calculations are sensitive to
even small variations in partial charges, especially when involving
highly charged guests and strongly polarizable hosts like CB7. For
example, we compared our AM1-BCC charges to those from Taproom and
found the root-mean-square (RMS) difference to be $0.018e$, with
the maximum atomic deviation reaching $0.0945e$. For any GB model,
the cavity radii and partial charge models are key determinants of
accuracy in binding free energy calculations. Setiadi et al. emphasized
the benefits of cavity radii reparameterization for reducing electrostatic
imbalance. Our results show that careful selection of a charge model---such
as using RESP instead of AM1-BCC---can provide an effective and potentially
simpler route to mitigate over binding. These two strategies may be
viewed as complementary levers for improving the reliability of implicit
solvent calculations, particularly in systems where electrostatics
dominate host--guest interactions. The AM1-BCC ABFEs are generally
more negative than those from PyRED RESP charges, while results using
Taproom charges are even more negative, often 20 to 50 kcal/mol more
negative than those computed with PyRED RESP charges.

We note that subsets of the results for CB7 and OAH were examined
by participants in SAMPL4\cite{muddanaSAMPL4HostGuest2014b}, and
WP6 in SAMPL9\cite{amezcuaSAMPL9HostGuestBlind2023}. While our results
compare favorably, direct comparison is outside the scope of this
study and would not be strictly fair, as participants submitted blind
predictions to the SAMPL challenges, whereas our results were obtained
retrospectively. Instead, our results demonstrate the utility of the
approach, which could yield even better agreement with experiment
if paired with an improved implicit solvation model.

A possible approach to overcome the limitations of the GB models tested
here is to apply bookending, in which the transfer free energies of
the bound and unbound states into a more accurate solvent model are
combined with the original result from \code{ISDDM.py} to yield corrected
ABFEs. Implicit solvation models that better captures the physics
of solvation--such as the numerical surface generalized Born model
with R6 integration (GBNSR6) \cite{aguilar_reducing_2010,forouzesh_grid-based_2017},
analytical generalized Born plus non-polar 2 model (AGBNP2) \cite{gallicchio_agbnp2_2009},
or three-dimensional reference interaction site models (3D-RISM) \cite{kovalenko1999selfconsistent}--can
serve as more physically realistic endpoints. GBNSR6 and AGBNP2 are
grid-based methods that may better represent Poisson--Boltzmann electrostatics,
introduce additional terms to account for solute--solvent dispersion,
hydration shell structuring, and specific solute hydration sites.
Alternatively, 3D-RISM uses a statistical mechanical approach to calculate
solvent structure and solute--solvent correlations with explicit
solvent models, naturally incorporating key phenomena such as charge
hydration asymmetry \cite{mukhopadhyay_charge_2012,mukhopadhyay_introducing_2014},
ion-specific effects\cite{giambasu2014ioncounting,giambasu2015competitive},
and solvent packing\cite{nguyen2019amolecular}. Incorporating these
models via bookending may provide more physically grounded and transferable
predictions for binding affinities and hydration free energies, mitigating
the limitations of traditional generalized Born radii while leveraging
them for sampling.

\begin{table*}
\begin{tabular}{ccccc}
\toprule 
System & AM1-BCC & RESP & Taproom & Unoptimized from Ref. \cite{setiadi2024tuningpotential}\tabularnewline
\midrule
cb7\_adz & -27.80 & -23.30 & -41.86 & -45.66\tabularnewline
cb7\_axm & -22.11 & -13.51 & -50.07 & -62.37\tabularnewline
cb7\_cha & -31.83 & -13.35 & -52.26 & -68.49\tabularnewline
cb7\_hxm & -18.15 & -11.53 & -34.98 & -44.36\tabularnewline
\bottomrule
\end{tabular}\caption{Comparison of CB7 binding free energies across charge models and force
field variants.\label{tab:Comparison-of-charge-models}}
\end{table*}

\subsection{Costs and Benefits of methodology}

Many methods have been develop to predict ABFEs and RBFEs, and our
conformationally restrained DDM workflow combines many of the benefits
of these methods. The primary benefit is that modified DDM in GB implicit
solvent is much faster than in the traditional application in explicit
solvent due to the use of an implicit solvent model. It is not only
faster per thermodynamic step, but it requires fewer intermediate
states to complete the cycle. Furthermore, it simplifies the procedure
by eliminating some of the most technically complicated steps in alchemical-pathway
methods, traditional DDM included.

A principal benefit is that the use of conformational restraints with
implicit solvent models is that the ligand vdW, Coulomb and solvent
interactions can each be turned on or off in a single step, removing
a major source of complexity present in many free energy methods \cite{klimovich_guidelines_2015,naden2015linearbasis}.
This eliminates the need for soft-core LJ potentials, which are commonly
used in alchemical-pathway methods to avoid the ``end-point catastrophe''
that occurs with steric overlaps as ligand is created or annihilated\cite{beutler_avoiding_1994}.
This steric overlap occurs between explicit solvent and the ligand,
but may also occur with the receptor as the ligand explores conformations
in the binding site. The use of soft-core potentials complicates BFE
calculations by adding significant complexity, requiring special code
to handle parts of the calculation \cite{naden2015linearbasis}, and
requiring additional resources to to simulation and process the intermediate
states with the MBAR or BAR methods \cite{li_repulsive_2020}. Much
work has been done to attempt to simplify or remove the use of soft-core
potentials due to the additional cost of calculation they entail\cite{lee_improved_2020,li_repulsive_2020}.
Through the use of implicit solvent and conformational restraints,
we avoid the need for soft-core LJ potentials or complex protocols
and most of the complexity they bring.

This approach may be used with a wide variety of MD engines, as soft-core
potentials, sophisticated restraints and thermodynamic derivatives
have been avoided. The only requirements beyond basic molecular dynamics
are an implicit solvent model, harmonic restraints, and the ability
to turn off vdW between receptor and ligand atoms. The requirement
to selectively turn off receptor-ligand vdW interactions could also
be avoided by setting the ligand LJ parameters to zero, as we did
with charges.

Unlike many other methods that employ implicit solvent models, the
method also implements rigorous ABFE techniques usually employed only
in explicit solvent. This makes it potentially more suitable for wide-ranging
application across chemical space than some end-state methods such
as MM/GBSA or MM/PBSA.

While we have avoided technical complexities for both the user and
the MD engine, we have introduced new lambda windows for the conformational
restraints. Users still must ensure sufficient phase-space overlap
throughout the cycle, though our implementation of the workflow generates
this analysis automatically, and the procedure for adding conformational-restraint
windows is straightforward. Overall, we anticipate that the number
of intermediate-state simulations will be about half of what is needed
for a DDM calculation with explicit solvent. The computational saving
from reducing the number of intermediate simulations and using implicit
solvent will be explored in future work.

A final consideration is the cost and accuracy of the implicit solvent
model itself. We have discussed the accuracy of solvent models in
section \ref{subsec:Solvation-models}, but it is worth noting that
GB implementations, such as those in AmberTools, scale as $O\left(N^{2}\right)$,
with the number of atoms, $N$, while particle mesh Ewald calculations
used for explicit solvent scale as $O\left(N\,\ln N\right)$. As a
result, for a large enough system, explicit solvent simulations will
be faster\cite{anandakrishnan_speed_2015}. However, GB models allow
more efficient conformational sampling per step \cite{anandakrishnan_speed_2015},
so the comparison not simply a matter of comparing nanoseconds-per-day
of simulation time.

\section{Conclusions\label{sec:Conclusions}}

We developed a modified ABFE protocol that combines DDM, GB implicit
solvent models, and conformational restraints, and implemented it
in an automated, parallel workflow,\code{ISDDM.py}. This approach
was tested on 93 host--guest complexes using five different GB models
from AmberTools.

The method is robust and reproducible. Conformational restraints eliminate
the need for soft-core potentials, simplify the decoupling process,
and reduce sampling requirements in intermediate states. The use of
implicit solvent avoids periodic boundary artifacts and allows efficient
sampling of end states. The automated workflow handles system setup,
replica exchange simulations, and free energy analysis with minimal
user intervention. Across systems and simulation variants, the method
consistently produced converged results with low numerical uncertainty
and was insensitive to protocol variations, provided sufficient overlap
was maintained.

However, the accuracy of the binding free energy predictions was highly
sensitive to the force field parameters---particularly the GB radii
and partial atomic charges. Systematic errors were observed for ligands
containing ammonium and carboxylate groups, which were under- or over-stabilized,
respectively. In targeted comparisons, switching from AM1-BCC to RESP
charges shifted predicted binding free energies by as much as 40\,kcal/mol
for some complexes, demonstrating the critical importance of parameter
selection when using implicit solvent models.

Overall, the conformationally restrained GB/DDM framework implemented
in\code{ISDDM.py} offers a practical and efficient approach for ABFE
calculations, while highlighting the central role of parameterization
in predictive accuracy. Future work will explore application to protein-ligand
systems, coupling of this method with more detailed solvation models,
such as 3D-RISM, and extending the automation to include adaptive
lambda placement and parameter tuning.
\begin{acknowledgement}
This material is based upon work supported by the National Science
Foundation (NSF) under Grants CHE-2102668, CHE-2018427, CHE-2320718
and MRI-2320846.
\end{acknowledgement}
\begin{suppinfo}
Simulation input package (.zip). Input scripts and starting structures
for the AMBER simulations described in this work are provided in the
accompanying compressed archive. These include topology/coordinate
files, restraint definitions, and example run scripts, allowing full
reproduction of the calculations. 

Free energy data (CSV). Two CSV files are provided:
\begin{itemize}
\item GB\_ABFE\_uncorrected.csv -- Calculated binding free energies (\textgreek{\textDelta}G)
from the GB workflow, without bias correction.
\item GB\_ABFE\_corrected.csv -- Same dataset with linear parameter correction
applied.
\end{itemize}
Both files contain the following columns:
\begin{itemize}
\item id -- Unique identifier for each host--guest system.
\item calculated\_deltaG -- Binding free energy (\textgreek{\textDelta}G,
kcal/mol) obtained from the GB workflow.
\item experimental\_deltaG -- Reference experimental binding free energy
(\textgreek{\textDelta}G, kcal/mol).
\item formal\_charges -- Net formal charge of the guest molecule.
\item model -- Implicit solvent model used in the calculation.
\item host -- Host receptor.
\item guest -- Guest identifier.
\end{itemize}
Figure S1: Host--guest systems based on cucurbit{[}7{]}uril (CB7)
and octa-acid (OA). The figure shows CB7 with 14 charged guest ligands
and OA with 23 guest ligands.

Figure S2. Host--guest systems based on \textgreek{\textalpha}-cyclodextrin
(ACD) and \textgreek{\textbeta}-cyclodextrin (BCD). 

Figure S3. Host--guest systems based on pillar{[}6{]}arene (WP6)
and 13 guests. 

Figure S3. Overlap matrix for the CB7--adn complex in the OBC model.
\end{suppinfo}

\bibliography{manuscript}

\providecommand{\latin}[1]{#1}
\makeatletter
\providecommand{\doi}
  {\begingroup\let\do\@makeother\dospecials
  \catcode`\{=1 \catcode`\}=2 \doi@aux}
\providecommand{\doi@aux}[1]{\endgroup\texttt{#1}}
\makeatother
\providecommand*\mcitethebibliography{\thebibliography}
\csname @ifundefined\endcsname{endmcitethebibliography}
  {\let\endmcitethebibliography\endthebibliography}{}
\begin{mcitethebibliography}{81}
\providecommand*\natexlab[1]{#1}
\providecommand*\mciteSetBstSublistMode[1]{}
\providecommand*\mciteSetBstMaxWidthForm[2]{}
\providecommand*\mciteBstWouldAddEndPuncttrue
  {\def\EndOfBibitem{\unskip.}}
\providecommand*\mciteBstWouldAddEndPunctfalse
  {\let\EndOfBibitem\relax}
\providecommand*\mciteSetBstMidEndSepPunct[3]{}
\providecommand*\mciteSetBstSublistLabelBeginEnd[3]{}
\providecommand*\EndOfBibitem{}
\mciteSetBstSublistMode{f}
\mciteSetBstMaxWidthForm{subitem}{(\alph{mcitesubitemcount})}
\mciteSetBstSublistLabelBeginEnd
  {\mcitemaxwidthsubitemform\space}
  {\relax}
  {\relax}

\bibitem[Deng and Roux(2009)Deng, and
  Roux]{dengComputationsStandardBinding2009a}
Deng,~Y.; Roux,~B. Computations of {{Standard Binding Free Energies}} with
  {{Molecular Dynamics Simulations}}. \emph{The Journal of Physical Chemistry
  B} \textbf{2009}, \emph{113}, 2234--2246\relax
\mciteBstWouldAddEndPuncttrue
\mciteSetBstMidEndSepPunct{\mcitedefaultmidpunct}
{\mcitedefaultendpunct}{\mcitedefaultseppunct}\relax
\EndOfBibitem
\bibitem[Ganesan \latin{et~al.}(2017)Ganesan, Coote, and
  Barakat]{ganesan2017molecular}
Ganesan,~A.; Coote,~M.~L.; Barakat,~K. Molecular dynamics-driven drug
  discovery: leaping forward with confidence. \emph{Drug Discovery Today}
  \textbf{2017}, \emph{22}, 249--269\relax
\mciteBstWouldAddEndPuncttrue
\mciteSetBstMidEndSepPunct{\mcitedefaultmidpunct}
{\mcitedefaultendpunct}{\mcitedefaultseppunct}\relax
\EndOfBibitem
\bibitem[Salo-Ahen \latin{et~al.}(2021)Salo-Ahen, Alanko, Bhadane, Bonvin,
  Honorato, Hossain, Juffer, Kabedev, Lahtela-Kakkonen, Larsen, Lescrinier,
  Marimuthu, Mirza, Mustafa, Nunes-Alves, Pantsar, Saadabadi, Singaravelu, and
  Vanmeert]{salo-ahen2021molecular}
Salo-Ahen,~O. M.~H. \latin{et~al.}  Molecular {Dynamics} {Simulations} in
  {Drug} {Discovery} and {Pharmaceutical} {Development}. \emph{Processes}
  \textbf{2021}, \emph{9}, 71\relax
\mciteBstWouldAddEndPuncttrue
\mciteSetBstMidEndSepPunct{\mcitedefaultmidpunct}
{\mcitedefaultendpunct}{\mcitedefaultseppunct}\relax
\EndOfBibitem
\bibitem[Hollingsworth and Dror(2018)Hollingsworth, and
  Dror]{hollingsworth2018molecular}
Hollingsworth,~S.~A.; Dror,~R.~O. Molecular {Dynamics} {Simulation} for {All}.
  \emph{Neuron} \textbf{2018}, \emph{99}, 1129--1143\relax
\mciteBstWouldAddEndPuncttrue
\mciteSetBstMidEndSepPunct{\mcitedefaultmidpunct}
{\mcitedefaultendpunct}{\mcitedefaultseppunct}\relax
\EndOfBibitem
\bibitem[Muddana \latin{et~al.}(2014)Muddana, Fenley, Mobley, and
  Gilson]{muddanaSAMPL4HostGuest2014b}
Muddana,~H.~S.; Fenley,~A.~T.; Mobley,~D.~L.; Gilson,~M.~K. The {{SAMPL4}}
  host-guest blind prediction challenge: an overview. \emph{Journal of
  Computer-Aided Molecular Design} \textbf{2014}, \emph{28}, 305--317\relax
\mciteBstWouldAddEndPuncttrue
\mciteSetBstMidEndSepPunct{\mcitedefaultmidpunct}
{\mcitedefaultendpunct}{\mcitedefaultseppunct}\relax
\EndOfBibitem
\bibitem[Heinzelmann and Gilson(2020)Heinzelmann, and
  Gilson]{heinzelmannAutomatedDockingRefinement2020a}
Heinzelmann,~G.; Gilson,~M.~K. \emph{Automated docking refinement and virtual
  compound screening with absolute binding free energy calculations}; Preprint,
  2020; p 2020.04.15.043240\relax
\mciteBstWouldAddEndPuncttrue
\mciteSetBstMidEndSepPunct{\mcitedefaultmidpunct}
{\mcitedefaultendpunct}{\mcitedefaultseppunct}\relax
\EndOfBibitem
\bibitem[Schindler \latin{et~al.}(2020)Schindler, Baumann, Blum, B{\"o}se,
  Buchstaller, Burgdorf, Cappel, Chekler, Czodrowski, Dorsch, Eguida, Follows,
  Fuch{\ss}, Gr{\"a}dler, Gunera, Johnson, Jorand~Lebrun, Karra, Klein,
  Knehans, Koetzner, Krier, Leiendecker, Leuthner, Li, Mochalkin, Musil, Neagu,
  Rippmann, Schiemann, Schulz, Steinbrecher, Tanzer, Unzue~Lopez,
  Viacava~Follis, Wegener, and Kuhn]{schindlerLargeScaleAssessmentBinding2020a}
Schindler,~C. E.~M. \latin{et~al.}  Large-{{Scale Assessment}} of {{Binding
  Free Energy Calculations}} in {{Active Drug Discovery Projects}}.
  \emph{Journal of Chemical Information and Modeling} \textbf{2020}, \emph{60},
  5457--5474\relax
\mciteBstWouldAddEndPuncttrue
\mciteSetBstMidEndSepPunct{\mcitedefaultmidpunct}
{\mcitedefaultendpunct}{\mcitedefaultseppunct}\relax
\EndOfBibitem
\bibitem[Cournia \latin{et~al.}(2020)Cournia, Allen, Beuming, Pearlman, Radak,
  and Sherman]{courniaRigorousFreeEnergy2020b}
Cournia,~Z.; Allen,~B.~K.; Beuming,~T.; Pearlman,~D.~A.; Radak,~B.~K.;
  Sherman,~W. Rigorous {{Free Energy Simulations}} in {{Virtual Screening}}.
  \emph{Journal of Chemical Information and Modeling} \textbf{2020}, \emph{60},
  4153--4169\relax
\mciteBstWouldAddEndPuncttrue
\mciteSetBstMidEndSepPunct{\mcitedefaultmidpunct}
{\mcitedefaultendpunct}{\mcitedefaultseppunct}\relax
\EndOfBibitem
\bibitem[Hou \latin{et~al.}(2011)Hou, Wang, Li, and Wang]{hou2011assessing}
Hou,~T.; Wang,~J.; Li,~Y.; Wang,~W. Assessing the {Performance} of the
  {MM}/{PBSA} and {MM}/{GBSA} {Methods}. 1. {The} {Accuracy} of {Binding}
  {Free} {Energy} {Calculations} {Based} on {Molecular} {Dynamics}
  {Simulations}. \emph{Journal of Chemical Information and Modeling}
  \textbf{2011}, \emph{51}, 69--82\relax
\mciteBstWouldAddEndPuncttrue
\mciteSetBstMidEndSepPunct{\mcitedefaultmidpunct}
{\mcitedefaultendpunct}{\mcitedefaultseppunct}\relax
\EndOfBibitem
\bibitem[Miller \latin{et~al.}(2012)Miller, McGee, Swails, Homeyer, Gohlke, and
  Roitberg]{miller2012mmpbsapy}
Miller,~B.~R.; McGee,~T.~D.; Swails,~J.~M.; Homeyer,~N.; Gohlke,~H.;
  Roitberg,~A.~E. {MMPBSA}.py: {An} {Efficient} {Program} for {End}-{State}
  {Free} {Energy} {Calculations}. \emph{Journal of Chemical Theory and
  Computation} \textbf{2012}, \emph{8}, 3314--3321\relax
\mciteBstWouldAddEndPuncttrue
\mciteSetBstMidEndSepPunct{\mcitedefaultmidpunct}
{\mcitedefaultendpunct}{\mcitedefaultseppunct}\relax
\EndOfBibitem
\bibitem[Srinivasan \latin{et~al.}(1998)Srinivasan, Cheatham, Cieplak, Kollman,
  and Case]{srinivasan1998continuum}
Srinivasan,~J.; Cheatham,~T.~E.; Cieplak,~P.; Kollman,~P.~A.; Case,~D.~A.
  Continuum solvent studies of the stability of {DNA}, {RNA}, and
  phosphoramidate - {DNA} helices. \emph{Journal of the American Chemical
  Society} \textbf{1998}, \emph{120}, 9401--9409\relax
\mciteBstWouldAddEndPuncttrue
\mciteSetBstMidEndSepPunct{\mcitedefaultmidpunct}
{\mcitedefaultendpunct}{\mcitedefaultseppunct}\relax
\EndOfBibitem
\bibitem[{\r A}qvist \latin{et~al.}(1994){\r A}qvist, Medina, and
  Samuelsson]{raqvist1994anew}
{\r A}qvist,~J.; Medina,~C.; Samuelsson,~J.-E. A new method for predicting
  binding affinity in computer-aided drug design. \emph{Protein Engineering,
  Design and Selection} \textbf{1994}, \emph{7}, 385--391\relax
\mciteBstWouldAddEndPuncttrue
\mciteSetBstMidEndSepPunct{\mcitedefaultmidpunct}
{\mcitedefaultendpunct}{\mcitedefaultseppunct}\relax
\EndOfBibitem
\bibitem[Lee \latin{et~al.}(1992)Lee, Chu, Bolger, and
  Warshel]{lee1992calculations}
Lee,~F.~S.; Chu,~Z.-T.; Bolger,~M.~B.; Warshel,~A. Calculations of
  antibody-antigen interactions: microscopic and semi-microscopic evaluation of
  the free energies of binding of phosphorylcholine analogs to {McPC603}.
  \emph{Protein Engineering, Design and Selection} \textbf{1992}, \emph{5},
  215--228\relax
\mciteBstWouldAddEndPuncttrue
\mciteSetBstMidEndSepPunct{\mcitedefaultmidpunct}
{\mcitedefaultendpunct}{\mcitedefaultseppunct}\relax
\EndOfBibitem
\bibitem[Henriksen \latin{et~al.}(2015)Henriksen, Fenley, and
  Gilson]{henriksen2015computational}
Henriksen,~N.~M.; Fenley,~A.~T.; Gilson,~M.~K. Computational calorimetry:
  high-precision calculation of host-guest binding thermodynamics.
  \emph{Journal of Chemical Theory and Computation} \textbf{2015}, \emph{11},
  4377--4394\relax
\mciteBstWouldAddEndPuncttrue
\mciteSetBstMidEndSepPunct{\mcitedefaultmidpunct}
{\mcitedefaultendpunct}{\mcitedefaultseppunct}\relax
\EndOfBibitem
\bibitem[Heinzelmann \latin{et~al.}(2017)Heinzelmann, Henriksen, and
  Gilson]{heinzelmann2017attachpullrelease}
Heinzelmann,~G.; Henriksen,~N.~M.; Gilson,~M.~K. Attach-{Pull}-{Release}
  {Calculations} of {Ligand} {Binding} and {Conformational} {Changes} on the
  {First} {BRD4} {Bromodomain}. \emph{Journal of Chemical Theory and
  Computation} \textbf{2017}, \emph{13}, 3260--3275\relax
\mciteBstWouldAddEndPuncttrue
\mciteSetBstMidEndSepPunct{\mcitedefaultmidpunct}
{\mcitedefaultendpunct}{\mcitedefaultseppunct}\relax
\EndOfBibitem
\bibitem[Kilburg and Gallicchio(2018)Kilburg, and
  Gallicchio]{kilburg2018assessment}
Kilburg,~D.; Gallicchio,~E. Assessment of a {Single} {Decoupling} {Alchemical}
  {Approach} for the {Calculation} of the {Absolute} {Binding} {Free}
  {Energies} of {Protein}-{Peptide} {Complexes}. \emph{Frontiers in Molecular
  Biosciences} \textbf{2018}, \emph{5}\relax
\mciteBstWouldAddEndPuncttrue
\mciteSetBstMidEndSepPunct{\mcitedefaultmidpunct}
{\mcitedefaultendpunct}{\mcitedefaultseppunct}\relax
\EndOfBibitem
\bibitem[Sakae \latin{et~al.}(2020)Sakae, Zhang, Levy, and
  Deng]{sakae2020absolute}
Sakae,~Y.; Zhang,~B.~W.; Levy,~R.~M.; Deng,~N. Absolute {Protein} {Binding}
  {Free} {Energy} {Simulations} for {Ligands} with {Multiple} {Poses}, a
  {Thermodynamic} {Path} {That} {Avoids} {Exhaustive} {Enumeration} of the
  {Poses}. \emph{Journal of Computational Chemistry} \textbf{2020}, \emph{41},
  56--68\relax
\mciteBstWouldAddEndPuncttrue
\mciteSetBstMidEndSepPunct{\mcitedefaultmidpunct}
{\mcitedefaultendpunct}{\mcitedefaultseppunct}\relax
\EndOfBibitem
\bibitem[Boresch \latin{et~al.}(2003)Boresch, Tettinger, Leitgeb, and
  Karplus]{boresch2003absolute}
Boresch,~S.; Tettinger,~F.; Leitgeb,~M.; Karplus,~M. Absolute {Binding} {Free}
  {Energies}: {A} {Quantitative} {Approach} for {Their} {Calculation}.
  \emph{The Journal of Physical Chemistry B} \textbf{2003}, \emph{107},
  9535--9551\relax
\mciteBstWouldAddEndPuncttrue
\mciteSetBstMidEndSepPunct{\mcitedefaultmidpunct}
{\mcitedefaultendpunct}{\mcitedefaultseppunct}\relax
\EndOfBibitem
\bibitem[Gilson \latin{et~al.}(1997)Gilson, Given, Bush, and
  McCammon]{gilsonStatisticalthermodynamicBasisComputation1997b}
Gilson,~M.~K.; Given,~J.~A.; Bush,~B.~L.; McCammon,~J.~A. The
  statistical-thermodynamic basis for computation of binding affinities: a
  critical review. \emph{Biophysical Journal} \textbf{1997}, \emph{72},
  1047--1069\relax
\mciteBstWouldAddEndPuncttrue
\mciteSetBstMidEndSepPunct{\mcitedefaultmidpunct}
{\mcitedefaultendpunct}{\mcitedefaultseppunct}\relax
\EndOfBibitem
\bibitem[Durrant and McCammon(2011)Durrant, and
  McCammon]{durrantMolecularDynamicsSimulations2011a}
Durrant,~J.~D.; McCammon,~J.~A. Molecular dynamics simulations and drug
  discovery. \emph{BMC Biology} \textbf{2011}, \emph{9}, 71\relax
\mciteBstWouldAddEndPuncttrue
\mciteSetBstMidEndSepPunct{\mcitedefaultmidpunct}
{\mcitedefaultendpunct}{\mcitedefaultseppunct}\relax
\EndOfBibitem
\bibitem[Anandakrishnan \latin{et~al.}(2015)Anandakrishnan, Drozdetski, Walker,
  and Onufriev]{anandakrishnan_speed_2015}
Anandakrishnan,~R.; Drozdetski,~A.; Walker,~R.; Onufriev,~A. Speed of
  conformational change: comparing explicit and implicit solvent molecular
  dynamics simulations. \emph{Biophysical Journal} \textbf{2015}, \emph{108},
  1153--1164\relax
\mciteBstWouldAddEndPuncttrue
\mciteSetBstMidEndSepPunct{\mcitedefaultmidpunct}
{\mcitedefaultendpunct}{\mcitedefaultseppunct}\relax
\EndOfBibitem
\bibitem[Khalak \latin{et~al.}(2021)Khalak, Tresdern, Aldeghi, Baumann, Mobley,
  Groot, and Gapsys]{khalak2021alchemical}
Khalak,~Y.; Tresdern,~G.; Aldeghi,~M.; Baumann,~H.~M.; Mobley,~D.~L.;
  Groot,~B.~d.; Gapsys,~V. Alchemical {Absolute} {Protein}-{Ligand} {Binding}
  {Free} {Energies} for {Drug} {Design}. \emph{Chemical Science} \textbf{2021},
  \relax
\mciteBstWouldAddEndPunctfalse
\mciteSetBstMidEndSepPunct{\mcitedefaultmidpunct}
{}{\mcitedefaultseppunct}\relax
\EndOfBibitem
\bibitem[Aldeghi \latin{et~al.}(2017)Aldeghi, Heifetz, Bodkin, Knapp, and
  Biggin]{aldeghi2017predictions}
Aldeghi,~M.; Heifetz,~A.; Bodkin,~M.~J.; Knapp,~S.; Biggin,~P.~C. Predictions
  of {Ligand} {Selectivity} from {Absolute} {Binding} {Free} {Energy}
  {Calculations}. \emph{Journal of the American Chemical Society}
  \textbf{2017}, \emph{139}, 946--957\relax
\mciteBstWouldAddEndPuncttrue
\mciteSetBstMidEndSepPunct{\mcitedefaultmidpunct}
{\mcitedefaultendpunct}{\mcitedefaultseppunct}\relax
\EndOfBibitem
\bibitem[Darve \latin{et~al.}(2008)Darve, Rodr{\'i}guez-G{\'o}mez, and
  Pohorille]{darve2008adaptive}
Darve,~E.; Rodr{\'i}guez-G{\'o}mez,~D.; Pohorille,~A. Adaptive biasing force
  method for scalar and vector free energy calculations. \emph{The Journal of
  Chemical Physics} \textbf{2008}, \emph{128}, 144120\relax
\mciteBstWouldAddEndPuncttrue
\mciteSetBstMidEndSepPunct{\mcitedefaultmidpunct}
{\mcitedefaultendpunct}{\mcitedefaultseppunct}\relax
\EndOfBibitem
\bibitem[Patey and Valleau(1975)Patey, and Valleau]{patey1975amonte}
Patey,~G.~N.; Valleau,~J.~P. A {Monte} {Carlo} method for obtaining the
  interionic potential of mean force in ionic solution. \emph{The Journal of
  Chemical Physics} \textbf{1975}, \emph{63}, 2334--2339\relax
\mciteBstWouldAddEndPuncttrue
\mciteSetBstMidEndSepPunct{\mcitedefaultmidpunct}
{\mcitedefaultendpunct}{\mcitedefaultseppunct}\relax
\EndOfBibitem
\bibitem[Izrailev \latin{et~al.}(1999)Izrailev, Stepaniants, Isralewitz,
  Kosztin, Lu, Molnar, Wriggers, and Schulten]{griebel_steered_1999}
Izrailev,~S.; Stepaniants,~S.; Isralewitz,~B.; Kosztin,~D.; Lu,~H.; Molnar,~F.;
  Wriggers,~W.; Schulten,~K. Steered molecular dynamics. Computational
  Molecular Dynamics: Challenges, Methods, Ideas. Berlin, Heidelberg, 1999; pp
  39--65\relax
\mciteBstWouldAddEndPuncttrue
\mciteSetBstMidEndSepPunct{\mcitedefaultmidpunct}
{\mcitedefaultendpunct}{\mcitedefaultseppunct}\relax
\EndOfBibitem
\bibitem[Laio and Parrinello(2002)Laio, and Parrinello]{laio_escaping_2002}
Laio,~A.; Parrinello,~M. Escaping free-energy minima. \emph{Proceedings of the
  National Academy of Sciences} \textbf{2002}, \emph{99}, 12562--12566\relax
\mciteBstWouldAddEndPuncttrue
\mciteSetBstMidEndSepPunct{\mcitedefaultmidpunct}
{\mcitedefaultendpunct}{\mcitedefaultseppunct}\relax
\EndOfBibitem
\bibitem[Sugita and Okamoto(1999)Sugita, and
  Okamoto]{sugita1999replicaexchange}
Sugita,~Y.; Okamoto,~Y. Replica-exchange molecular dynamics method for protein
  folding. \emph{Chemical Physics Letters} \textbf{1999}, \emph{314},
  141--151\relax
\mciteBstWouldAddEndPuncttrue
\mciteSetBstMidEndSepPunct{\mcitedefaultmidpunct}
{\mcitedefaultendpunct}{\mcitedefaultseppunct}\relax
\EndOfBibitem
\bibitem[Onuchic and Wolynes(2004)Onuchic, and Wolynes]{onuchic2004theoryof}
Onuchic,~J.~N.; Wolynes,~P.~G. Theory of protein folding. \emph{Current Opinion
  in Structural Biology} \textbf{2004}, \emph{14}, 70--75\relax
\mciteBstWouldAddEndPuncttrue
\mciteSetBstMidEndSepPunct{\mcitedefaultmidpunct}
{\mcitedefaultendpunct}{\mcitedefaultseppunct}\relax
\EndOfBibitem
\bibitem[Deng \latin{et~al.}(2018)Deng, Cui, Zhang, Xia, Cruz, and
  Levy]{deng_comparing_2018}
Deng,~N.; Cui,~D.; Zhang,~B.~W.; Xia,~J.; Cruz,~J.; Levy,~R. Comparing
  alchemical and physical pathway methods for computing the absolute binding
  free energy of charged ligands. \emph{Physical Chemistry Chemical Physics}
  \textbf{2018}, \emph{20}, 17081--17092\relax
\mciteBstWouldAddEndPuncttrue
\mciteSetBstMidEndSepPunct{\mcitedefaultmidpunct}
{\mcitedefaultendpunct}{\mcitedefaultseppunct}\relax
\EndOfBibitem
\bibitem[{\"O}hlknecht \latin{et~al.}(2020){\"O}hlknecht, Lier, Petrov, Fuchs,
  and Oostenbrink]{ohlknecht2020correcting}
{\"O}hlknecht,~C.; Lier,~B.; Petrov,~D.; Fuchs,~J.; Oostenbrink,~C. Correcting
  electrostatic artifacts due to net-charge changes in the calculation of
  ligand binding free energies. \emph{Journal of Computational Chemistry}
  \textbf{2020}, \emph{41}, 986--999\relax
\mciteBstWouldAddEndPuncttrue
\mciteSetBstMidEndSepPunct{\mcitedefaultmidpunct}
{\mcitedefaultendpunct}{\mcitedefaultseppunct}\relax
\EndOfBibitem
\bibitem[Rocklin \latin{et~al.}(2013)Rocklin, Mobley, Dill, and
  H{\"u}nenberger]{rocklin2013calculating}
Rocklin,~G.~J.; Mobley,~D.~L.; Dill,~K.~A.; H{\"u}nenberger,~P.~H. Calculating
  the binding free energies of charged species based on explicit-solvent
  simulations employing lattice-sum methods: {An} accurate correction scheme
  for electrostatic finite-size effects. \emph{The Journal of Chemical Physics}
  \textbf{2013}, \emph{139}, 184103\relax
\mciteBstWouldAddEndPuncttrue
\mciteSetBstMidEndSepPunct{\mcitedefaultmidpunct}
{\mcitedefaultendpunct}{\mcitedefaultseppunct}\relax
\EndOfBibitem
\bibitem[Lee \latin{et~al.}(2020)Lee, Allen, Giese, Guo, Li, Lin, McGee,
  Pearlman, Radak, Tao, Tsai, Xu, Sherman, and
  York]{leeAlchemicalBindingFree2020a}
Lee,~T.-S.; Allen,~B.~K.; Giese,~T.~J.; Guo,~Z.; Li,~P.; Lin,~C.; McGee,~T.~D.;
  Pearlman,~D.~A.; Radak,~B.~K.; Tao,~Y.; Tsai,~H.-C.; Xu,~H.; Sherman,~W.;
  York,~D.~M. Alchemical {{Binding Free Energy Calculations}} in {{AMBER20}}:
  {{Advances}} and {{Best Practices}} for {{Drug Discovery}}. \emph{Journal of
  Chemical Information and Modeling} \textbf{2020}, \emph{60}, 5595--5623\relax
\mciteBstWouldAddEndPuncttrue
\mciteSetBstMidEndSepPunct{\mcitedefaultmidpunct}
{\mcitedefaultendpunct}{\mcitedefaultseppunct}\relax
\EndOfBibitem
\bibitem[Pohorille \latin{et~al.}(2010)Pohorille, Jarzynski, and
  Chipot]{pohorille2010goodpractices}
Pohorille,~A.; Jarzynski,~C.; Chipot,~C. Good {Practices} in {Free}-{Energy}
  {Calculations}. \emph{The Journal of Physical Chemistry B} \textbf{2010},
  \emph{114}, 10235--10253\relax
\mciteBstWouldAddEndPuncttrue
\mciteSetBstMidEndSepPunct{\mcitedefaultmidpunct}
{\mcitedefaultendpunct}{\mcitedefaultseppunct}\relax
\EndOfBibitem
\bibitem[Steinbrecher \latin{et~al.}(2007)Steinbrecher, Mobley, and
  Case]{steinbrecher2007nonlinear}
Steinbrecher,~T.; Mobley,~D.~L.; Case,~D.~A. Nonlinear scaling schemes for
  {Lennard}-{Jones} interactions in free energy calculations. \emph{The Journal
  of Chemical Physics} \textbf{2007}, \emph{127}, 214108\relax
\mciteBstWouldAddEndPuncttrue
\mciteSetBstMidEndSepPunct{\mcitedefaultmidpunct}
{\mcitedefaultendpunct}{\mcitedefaultseppunct}\relax
\EndOfBibitem
\bibitem[Beutler \latin{et~al.}(1994)Beutler, Mark, Van~Schaik, Gerber, and
  Van~Gunsteren]{beutler_avoiding_1994}
Beutler,~T.~C.; Mark,~A.~E.; Van~Schaik,~R.~C.; Gerber,~P.~R.;
  Van~Gunsteren,~W.~F. Avoiding singularities and numerical instabilities in
  free energy calculations based on molecular simulations. \emph{Chemical
  Physics Letters} \textbf{1994}, \emph{222}, 529--539\relax
\mciteBstWouldAddEndPuncttrue
\mciteSetBstMidEndSepPunct{\mcitedefaultmidpunct}
{\mcitedefaultendpunct}{\mcitedefaultseppunct}\relax
\EndOfBibitem
\bibitem[Still \latin{et~al.}(1990)Still, Tempczyk, Hawley, and
  Hendrickson]{still1990semianalytical}
Still,~W.~C.; Tempczyk,~A.; Hawley,~R.~C.; Hendrickson,~T. Semianalytical
  treatment of solvation for molecular mechanics and dynamics. \emph{Journal of
  the American Chemical Society} \textbf{1990}, \emph{112}, 6127--6129\relax
\mciteBstWouldAddEndPuncttrue
\mciteSetBstMidEndSepPunct{\mcitedefaultmidpunct}
{\mcitedefaultendpunct}{\mcitedefaultseppunct}\relax
\EndOfBibitem
\bibitem[Cruz \latin{et~al.}(2020)Cruz, Wickstrom, Yang, Gallicchio, and
  Deng]{cruz2020combining}
Cruz,~J.; Wickstrom,~L.; Yang,~D.; Gallicchio,~E.; Deng,~N. Combining
  {Alchemical} {Transformation} with a {Physical} {Pathway} to {Accelerate}
  {Absolute} {Binding} {Free} {Energy} {Calculations} of {Charged} {Ligands} to
  {Enclosed} {Binding} {Sites}. \emph{Journal of Chemical Theory and
  Computation} \textbf{2020}, \emph{16}, 2803--2813\relax
\mciteBstWouldAddEndPuncttrue
\mciteSetBstMidEndSepPunct{\mcitedefaultmidpunct}
{\mcitedefaultendpunct}{\mcitedefaultseppunct}\relax
\EndOfBibitem
\bibitem[Gallicchio \latin{et~al.}(2010)Gallicchio, Lapelosa, and
  Levy]{gallicchio2010binding}
Gallicchio,~E.; Lapelosa,~M.; Levy,~R.~M. Binding {Energy} {Distribution}
  {Analysis} {Method} ({BEDAM}) for {Estimation} of {Protein}-{Ligand}
  {Binding} {Affinities}. \emph{Journal of Chemical Theory and Computation}
  \textbf{2010}, \emph{6}, 2961--2977\relax
\mciteBstWouldAddEndPuncttrue
\mciteSetBstMidEndSepPunct{\mcitedefaultmidpunct}
{\mcitedefaultendpunct}{\mcitedefaultseppunct}\relax
\EndOfBibitem
\bibitem[Setiadi \latin{et~al.}(2024)Setiadi, Boothroyd, Slochower, Dotson,
  Thompson, Wagner, Wang, and Gilson]{setiadi2024tuningpotential}
Setiadi,~J.; Boothroyd,~S.; Slochower,~D.~R.; Dotson,~D.~L.; Thompson,~M.~W.;
  Wagner,~J.~R.; Wang,~L.-P.; Gilson,~M.~K. Tuning potential functions to
  host-binding data. \emph{Journal of Chemical Theory and Computation}
  \textbf{2024}, \emph{20}, 239--252\relax
\mciteBstWouldAddEndPuncttrue
\mciteSetBstMidEndSepPunct{\mcitedefaultmidpunct}
{\mcitedefaultendpunct}{\mcitedefaultseppunct}\relax
\EndOfBibitem
\bibitem[Case \latin{et~al.}(2023)Case, Aktulga, Belfon, Cerutti, Cisneros,
  Cruzeiro, Forouzesh, Giese, G{\"o}tz, Gohlke, Izadi, Kasavajhala, Kaymak,
  King, Kurtzman, Lee, Li, Liu, Luchko, Luo, Manathunga, Machado, Nguyen,
  O'Hearn, Onufriev, Pan, Pantano, Qi, Rahnamoun, Risheh, Schott-Verdugo,
  Shajan, Swails, Wang, Wei, Wu, Wu, Zhang, Zhao, Zhu, Cheatham, Roe, Roitberg,
  Simmerling, York, Nagan, and Merz]{case2023ambertools}
Case,~D.~A. \latin{et~al.}  {AmberTools}. \emph{Journal of Chemical Information
  and Modeling} \textbf{2023}, \relax
\mciteBstWouldAddEndPunctfalse
\mciteSetBstMidEndSepPunct{\mcitedefaultmidpunct}
{}{\mcitedefaultseppunct}\relax
\EndOfBibitem
\bibitem[Slochower(2022)]{slochowerTaproom2022}
Slochower,~D. Taproom. 2022\relax
\mciteBstWouldAddEndPuncttrue
\mciteSetBstMidEndSepPunct{\mcitedefaultmidpunct}
{\mcitedefaultendpunct}{\mcitedefaultseppunct}\relax
\EndOfBibitem
\bibitem[Roe and Cheatham(2013)Roe, and Cheatham]{roe_ptraj_2013}
Roe,~D.~R.; Cheatham,~T.~E. {PTRAJ} and {CPPTRAJ}: software for processing and
  analysis of molecular dynamics trajectory data. \emph{Journal of Chemical
  Theory and Computation} \textbf{2013}, \emph{9}, 3084--3095\relax
\mciteBstWouldAddEndPuncttrue
\mciteSetBstMidEndSepPunct{\mcitedefaultmidpunct}
{\mcitedefaultendpunct}{\mcitedefaultseppunct}\relax
\EndOfBibitem
\bibitem[Vivian \latin{et~al.}(2017)Vivian, Rao, Nothaft, Ketchum, Armstrong,
  Novak, Pfeil, Narkizian, Deran, Musselman-Brown, Schmidt, Amstutz, Craft,
  Goldman, Rosenbloom, Cline, O'Connor, Hanna, Birger, Kent, Patterson, Joseph,
  Zhu, Zaranek, Getz, Haussler, and Paten]{vivian_toil_2017}
Vivian,~J. \latin{et~al.}  Toil enables reproducible, open source, big
  biomedical data analyses. \emph{Nature Biotechnology} \textbf{2017},
  \emph{35}, 314--316\relax
\mciteBstWouldAddEndPuncttrue
\mciteSetBstMidEndSepPunct{\mcitedefaultmidpunct}
{\mcitedefaultendpunct}{\mcitedefaultseppunct}\relax
\EndOfBibitem
\bibitem[Shirts and Chodera(2008)Shirts, and Chodera]{shirts2008statistically}
Shirts,~M.~R.; Chodera,~J.~D. Statistically optimal analysis of samples from
  multiple equilibrium states. \emph{The Journal of Chemical Physics}
  \textbf{2008}, \emph{129}, 124105\relax
\mciteBstWouldAddEndPuncttrue
\mciteSetBstMidEndSepPunct{\mcitedefaultmidpunct}
{\mcitedefaultendpunct}{\mcitedefaultseppunct}\relax
\EndOfBibitem
\bibitem[Chodera(2016)]{chodera_simple_2016}
Chodera,~J.~D. A simple method for automated equilibration detection in
  molecular simulations. \emph{Journal of Chemical Theory and Computation}
  \textbf{2016}, \emph{12}, 1799--1805\relax
\mciteBstWouldAddEndPuncttrue
\mciteSetBstMidEndSepPunct{\mcitedefaultmidpunct}
{\mcitedefaultendpunct}{\mcitedefaultseppunct}\relax
\EndOfBibitem
\bibitem[Bennett(1976)]{bennettEfficientEstimationFree1976a}
Bennett,~C.~H. Efficient estimation of free energy differences from {{Monte
  Carlo}} data. \emph{Journal of Computational Physics} \textbf{1976},
  \emph{22}, 245--268\relax
\mciteBstWouldAddEndPuncttrue
\mciteSetBstMidEndSepPunct{\mcitedefaultmidpunct}
{\mcitedefaultendpunct}{\mcitedefaultseppunct}\relax
\EndOfBibitem
\bibitem[Klimovich \latin{et~al.}(2015)Klimovich, Shirts, and
  Mobley]{klimovich_guidelines_2015}
Klimovich,~P.~V.; Shirts,~M.~R.; Mobley,~D.~L. Guidelines for the analysis of
  free energy calculations. \emph{Journal of Computer-Aided Molecular Design}
  \textbf{2015}, \emph{29}, 397--411\relax
\mciteBstWouldAddEndPuncttrue
\mciteSetBstMidEndSepPunct{\mcitedefaultmidpunct}
{\mcitedefaultendpunct}{\mcitedefaultseppunct}\relax
\EndOfBibitem
\bibitem[Forli \latin{et~al.}(2016)Forli, Huey, Pique, Sanner, Goodsell, and
  Olson]{forli_computational_2016}
Forli,~S.; Huey,~R.; Pique,~M.~E.; Sanner,~M.~F.; Goodsell,~D.~S.; Olson,~A.~J.
  Computational protein-ligand docking and virtual drug screening with the
  {AutoDock} suite. \emph{Nature Protocols} \textbf{2016}, \emph{11},
  905--919\relax
\mciteBstWouldAddEndPuncttrue
\mciteSetBstMidEndSepPunct{\mcitedefaultmidpunct}
{\mcitedefaultendpunct}{\mcitedefaultseppunct}\relax
\EndOfBibitem
\bibitem[Jakalian \latin{et~al.}(2002)Jakalian, Jack, and
  Bayly]{jakalian_fast_2002}
Jakalian,~A.; Jack,~D.~B.; Bayly,~C.~I. Fast, efficient generation of high
  quality atomic charges. {AM}1 {BCC} model: {II}. Parameterization and
  validation. \emph{Journal of Computational Chemistry} \textbf{2002},
  \emph{23}, 1623--1641\relax
\mciteBstWouldAddEndPuncttrue
\mciteSetBstMidEndSepPunct{\mcitedefaultmidpunct}
{\mcitedefaultendpunct}{\mcitedefaultseppunct}\relax
\EndOfBibitem
\bibitem[Wang \latin{et~al.}(2004)Wang, Wolf, Caldwell, Kollman, and
  Case]{wang2004development}
Wang,~J.; Wolf,~R.~M.; Caldwell,~J.~W.; Kollman,~P.~A.; Case,~D.~A. Development
  and testing of a general amber force field. \emph{Journal of Computational
  Chemistry} \textbf{2004}, \emph{25}, 1157--1174\relax
\mciteBstWouldAddEndPuncttrue
\mciteSetBstMidEndSepPunct{\mcitedefaultmidpunct}
{\mcitedefaultendpunct}{\mcitedefaultseppunct}\relax
\EndOfBibitem
\bibitem[Cornell \latin{et~al.}(1995)Cornell, Cieplak, Bayly, Gould, Merz,
  Ferguson, Spellmeyer, Fox, Caldwell, and Kollman]{cornell_second_1995}
Cornell,~W.~D.; Cieplak,~P.; Bayly,~C.~I.; Gould,~I.~R.; Merz,~K.~M.;
  Ferguson,~D.~M.; Spellmeyer,~D.~C.; Fox,~T.; Caldwell,~J.~W.; Kollman,~P.~A.
  A second generation force field for the simulation of proteins, nucleic
  acids, and organic molecules. \emph{Journal of the American Chemical Society}
  \textbf{1995}, \emph{117}, 5179--5197\relax
\mciteBstWouldAddEndPuncttrue
\mciteSetBstMidEndSepPunct{\mcitedefaultmidpunct}
{\mcitedefaultendpunct}{\mcitedefaultseppunct}\relax
\EndOfBibitem
\bibitem[Hopkins \latin{et~al.}(2015)Hopkins, Le~Grand, Walker, and
  Roitberg]{hopkins_long-time-step_2015}
Hopkins,~C.~W.; Le~Grand,~S.; Walker,~R.~C.; Roitberg,~A.~E. Long-time-step
  molecular dynamics through hydrogen mass repartitioning. \emph{Journal of
  Chemical Theory and Computation} \textbf{2015}, \emph{11}, 1864--1874\relax
\mciteBstWouldAddEndPuncttrue
\mciteSetBstMidEndSepPunct{\mcitedefaultmidpunct}
{\mcitedefaultendpunct}{\mcitedefaultseppunct}\relax
\EndOfBibitem
\bibitem[Case \latin{et~al.}()Case, Ben-Shalom, Brozell, Cerutti, Cheatham,
  Cruzeiro, Duke, Darden, Ghoreishi, Giambasu, Giese, Gilson, Gohlke, Goetz,
  Greene, Harris, Homeyer, Huang, Izadi, Kovalenko, Krasny, Kurtzman, Lee,
  LeGrand, Li, Lin, Liu, Luchko, Luo, Man, Mermelstein, Merz, Miao, Monard,
  Nguyen, Nguyen, Onufriev, Pan, Qi, Roe, Roitberg, Sagui, Schott-Verdugo,
  Shen, Simmerling, Smith, Swails, Walker, Wang, Wei, Wilson, Wolf, Wu, Xiao,
  Xiong, York, and Kollman]{AMBER21}
Case,~D.~A. \latin{et~al.}  {AMBER} 21. \relax
\mciteBstWouldAddEndPunctfalse
\mciteSetBstMidEndSepPunct{\mcitedefaultmidpunct}
{}{\mcitedefaultseppunct}\relax
\EndOfBibitem
\bibitem[Hawkins \latin{et~al.}(1995)Hawkins, Cramer, and
  Truhlar]{hawkins_pairwise_1995}
Hawkins,~G.~D.; Cramer,~C.~J.; Truhlar,~D.~G. Pairwise solute descreening of
  solute charges from a dielectric medium. \emph{Chemical Physics Letters}
  \textbf{1995}, \emph{246}, 122--129\relax
\mciteBstWouldAddEndPuncttrue
\mciteSetBstMidEndSepPunct{\mcitedefaultmidpunct}
{\mcitedefaultendpunct}{\mcitedefaultseppunct}\relax
\EndOfBibitem
\bibitem[Onufriev \latin{et~al.}(2000)Onufriev, Bashford, and
  Case]{onufriev_modification_2000}
Onufriev,~A.; Bashford,~D.; Case,~D.~A. Modification of the generalized {Born}
  model suitable for macromolecules. \emph{The Journal of Physical Chemistry B}
  \textbf{2000}, \emph{104}, 3712--3720\relax
\mciteBstWouldAddEndPuncttrue
\mciteSetBstMidEndSepPunct{\mcitedefaultmidpunct}
{\mcitedefaultendpunct}{\mcitedefaultseppunct}\relax
\EndOfBibitem
\bibitem[Onufriev \latin{et~al.}(2004)Onufriev, Bashford, and
  Case]{onufriev_exploring_2004}
Onufriev,~A.; Bashford,~D.; Case,~D.~A. Exploring protein native states and
  large scale conformational changes with a modified generalized born model.
  \emph{Proteins: Structure, Function, and Bioinformatics} \textbf{2004},
  \emph{55}, 383--394\relax
\mciteBstWouldAddEndPuncttrue
\mciteSetBstMidEndSepPunct{\mcitedefaultmidpunct}
{\mcitedefaultendpunct}{\mcitedefaultseppunct}\relax
\EndOfBibitem
\bibitem[Mongan \latin{et~al.}(2007)Mongan, Simmerling, {McCammon}, Case, and
  Onufriev]{mongan_generalized_2007}
Mongan,~J.; Simmerling,~C.; {McCammon},~J.~A.; Case,~D.~A.; Onufriev,~A.
  Generalized {B}orn model with a simple, robust molecular volume correction.
  \emph{Journal of Chemical Theory and Computation} \textbf{2007}, \emph{3},
  156--169\relax
\mciteBstWouldAddEndPuncttrue
\mciteSetBstMidEndSepPunct{\mcitedefaultmidpunct}
{\mcitedefaultendpunct}{\mcitedefaultseppunct}\relax
\EndOfBibitem
\bibitem[Nguyen \latin{et~al.}(2013)Nguyen, Roe, and
  Simmerling]{nguyen2013improved}
Nguyen,~H.; Roe,~D.~R.; Simmerling,~C. Improved {Generalized} {Born} {Solvent}
  {Model} {Parameters} for {Protein} {Simulations}. \emph{Journal of chemical
  theory and computation} \textbf{2013}, \emph{9}, 2020--2034\relax
\mciteBstWouldAddEndPuncttrue
\mciteSetBstMidEndSepPunct{\mcitedefaultmidpunct}
{\mcitedefaultendpunct}{\mcitedefaultseppunct}\relax
\EndOfBibitem
\bibitem[Patriksson and Spoel(2008)Patriksson, and
  Spoel]{patriksson2008atemperature}
Patriksson,~A.; Spoel,~D. v.~d. A temperature predictor for parallel tempering
  simulations. \emph{Physical Chemistry Chemical Physics} \textbf{2008},
  \emph{10}, 2073--2077\relax
\mciteBstWouldAddEndPuncttrue
\mciteSetBstMidEndSepPunct{\mcitedefaultmidpunct}
{\mcitedefaultendpunct}{\mcitedefaultseppunct}\relax
\EndOfBibitem
\bibitem[Spoel(2023)]{spoel2023dspoelremdtemperaturegenerator}
Spoel,~D. v.~d. dspoel/remd-temperature-generator. 2023\relax
\mciteBstWouldAddEndPuncttrue
\mciteSetBstMidEndSepPunct{\mcitedefaultmidpunct}
{\mcitedefaultendpunct}{\mcitedefaultseppunct}\relax
\EndOfBibitem
\bibitem[Roe \latin{et~al.}(2014)Roe, Bergonzo, and
  Cheatham]{roe2014evaluation}
Roe,~D.~R.; Bergonzo,~C.; Cheatham,~T. E. I. I.~I. Evaluation of {Enhanced}
  {Sampling} {Provided} by {Accelerated} {Molecular} {Dynamics} with
  {Hamiltonian} {Replica} {Exchange} {Methods}. \emph{The Journal of Physical
  Chemistry B} \textbf{2014}, \emph{118}, 3543--3552\relax
\mciteBstWouldAddEndPuncttrue
\mciteSetBstMidEndSepPunct{\mcitedefaultmidpunct}
{\mcitedefaultendpunct}{\mcitedefaultseppunct}\relax
\EndOfBibitem
\bibitem[Onufriev and Case(2019)Onufriev, and Case]{onufriev2019generalized}
Onufriev,~A.~V.; Case,~D.~A. Generalized {Born} {Implicit} {Solvent} {Models}
  for {Biomolecules}. \emph{Annual Review of Biophysics} \textbf{2019},
  \emph{48}, 275--296\relax
\mciteBstWouldAddEndPuncttrue
\mciteSetBstMidEndSepPunct{\mcitedefaultmidpunct}
{\mcitedefaultendpunct}{\mcitedefaultseppunct}\relax
\EndOfBibitem
\bibitem[Bayly \latin{et~al.}(1993)Bayly, Cieplak, Cornell, and
  Kollman]{bayly_well-behaved_1993}
Bayly,~C.~I.; Cieplak,~P.; Cornell,~W.; Kollman,~P.~A. A well-behaved
  electrostatic potential based method using charge restraints for deriving
  atomic charges: the {RESP} model. \emph{The Journal of Physical Chemistry}
  \textbf{1993}, \emph{97}, 10269--10280\relax
\mciteBstWouldAddEndPuncttrue
\mciteSetBstMidEndSepPunct{\mcitedefaultmidpunct}
{\mcitedefaultendpunct}{\mcitedefaultseppunct}\relax
\EndOfBibitem
\bibitem[Vanquelef \latin{et~al.}(2011)Vanquelef, Simon, Marquant, Garcia,
  Klimerak, Delepine, Cieplak, and Dupradeau]{vanquelef_red_2011}
Vanquelef,~E.; Simon,~S.; Marquant,~G.; Garcia,~E.; Klimerak,~G.;
  Delepine,~J.~C.; Cieplak,~P.; Dupradeau,~F.-Y. {R.E.D. Server}: a web service
  for deriving {RESP} and {ESP} charges and building force field libraries for
  new molecules and molecular fragments. \emph{Nucleic Acids Research}
  \textbf{2011}, \emph{39}, W511--W517\relax
\mciteBstWouldAddEndPuncttrue
\mciteSetBstMidEndSepPunct{\mcitedefaultmidpunct}
{\mcitedefaultendpunct}{\mcitedefaultseppunct}\relax
\EndOfBibitem
\bibitem[He \latin{et~al.}(2020)He, Man, Yang, Lee, and Wang]{he_fast_2020}
He,~X.; Man,~V.~H.; Yang,~W.; Lee,~T.-S.; Wang,~J. A fast and high-quality
  charge model for the next generation general {AMBER} force field. \emph{The
  Journal of Chemical Physics} \textbf{2020}, \emph{153}, 114502\relax
\mciteBstWouldAddEndPuncttrue
\mciteSetBstMidEndSepPunct{\mcitedefaultmidpunct}
{\mcitedefaultendpunct}{\mcitedefaultseppunct}\relax
\EndOfBibitem
\bibitem[Velez~Vega and Gilson(2013)Velez~Vega, and
  Gilson]{velezvega_overcoming_2013}
Velez~Vega,~C.; Gilson,~M.~K. Overcoming dissipation in the calculation of
  standard binding free energies by ligand extraction. \emph{Journal of
  Computational Chemistry} \textbf{2013}, \emph{34}, 2360--2371\relax
\mciteBstWouldAddEndPuncttrue
\mciteSetBstMidEndSepPunct{\mcitedefaultmidpunct}
{\mcitedefaultendpunct}{\mcitedefaultseppunct}\relax
\EndOfBibitem
\bibitem[Amezcua \latin{et~al.}(2024)Amezcua, Setiadi, and
  Mobley]{amezcuaSAMPL9HostGuestBlind2023}
Amezcua,~M.; Setiadi,~J.; Mobley,~D.~L. The {SAMPL9} host-guest blind
  challenge: an overview of binding free energy predictive accuracy.
  \emph{Phys. Chem. Chem. Phys.} \textbf{2024}, \emph{26}, 9207--9225\relax
\mciteBstWouldAddEndPuncttrue
\mciteSetBstMidEndSepPunct{\mcitedefaultmidpunct}
{\mcitedefaultendpunct}{\mcitedefaultseppunct}\relax
\EndOfBibitem
\bibitem[Aguilar \latin{et~al.}(2010)Aguilar, Shadrach, and
  Onufriev]{aguilar_reducing_2010}
Aguilar,~B.; Shadrach,~R.; Onufriev,~A.~V. Reducing the secondary structure
  bias in the generalized {B}orn model via {R6} effective radii. \emph{Journal
  of Chemical Theory and Computation} \textbf{2010}, \emph{6}, 3613--3630\relax
\mciteBstWouldAddEndPuncttrue
\mciteSetBstMidEndSepPunct{\mcitedefaultmidpunct}
{\mcitedefaultendpunct}{\mcitedefaultseppunct}\relax
\EndOfBibitem
\bibitem[Forouzesh \latin{et~al.}(2017)Forouzesh, Izadi, and
  Onufriev]{forouzesh_grid-based_2017}
Forouzesh,~N.; Izadi,~S.; Onufriev,~A.~V. Grid-based surface generalized {B}orn
  model for calculation of electrostatic binding free energies. \emph{Journal
  of Chemical Information and Modeling} \textbf{2017}, \emph{57},
  2505--2513\relax
\mciteBstWouldAddEndPuncttrue
\mciteSetBstMidEndSepPunct{\mcitedefaultmidpunct}
{\mcitedefaultendpunct}{\mcitedefaultseppunct}\relax
\EndOfBibitem
\bibitem[Gallicchio \latin{et~al.}(2009)Gallicchio, Paris, and
  Levy]{gallicchio_agbnp2_2009}
Gallicchio,~E.; Paris,~K.; Levy,~R.~M. The {AGBNP}2 implicit solvation model.
  \emph{Journal of Chemical Theory and Computation} \textbf{2009}, \emph{5},
  2544--2564\relax
\mciteBstWouldAddEndPuncttrue
\mciteSetBstMidEndSepPunct{\mcitedefaultmidpunct}
{\mcitedefaultendpunct}{\mcitedefaultseppunct}\relax
\EndOfBibitem
\bibitem[Kovalenko and Hirata(1999)Kovalenko, and
  Hirata]{kovalenko1999selfconsistent}
Kovalenko,~A.; Hirata,~F. Self-consistent description of a metal-water
  interface by the {Kohn-Sham} density functional theory and the
  three-dimensional reference interaction site model. \emph{The Journal of
  Chemical Physics} \textbf{1999}, \emph{110}, 10095--10112\relax
\mciteBstWouldAddEndPuncttrue
\mciteSetBstMidEndSepPunct{\mcitedefaultmidpunct}
{\mcitedefaultendpunct}{\mcitedefaultseppunct}\relax
\EndOfBibitem
\bibitem[Mukhopadhyay \latin{et~al.}(2012)Mukhopadhyay, Fenley, Tolokh, and
  Onufriev]{mukhopadhyay_charge_2012}
Mukhopadhyay,~A.; Fenley,~A.~T.; Tolokh,~I.~S.; Onufriev,~A.~V. Charge
  hydration asymmetry: the basic principle and how to use it to test and
  improve water models. \emph{The Journal of Physical Chemistry B}
  \textbf{2012}, \emph{116}, 9776--9783\relax
\mciteBstWouldAddEndPuncttrue
\mciteSetBstMidEndSepPunct{\mcitedefaultmidpunct}
{\mcitedefaultendpunct}{\mcitedefaultseppunct}\relax
\EndOfBibitem
\bibitem[Mukhopadhyay \latin{et~al.}(2014)Mukhopadhyay, Aguilar, Tolokh, and
  Onufriev]{mukhopadhyay_introducing_2014}
Mukhopadhyay,~A.; Aguilar,~B.~H.; Tolokh,~I.~S.; Onufriev,~A.~V. Introducing
  charge hydration asymmetry into the generalized {B}orn model. \emph{Journal
  of Chemical Theory and Computation} \textbf{2014}, \emph{10},
  1788--1794\relax
\mciteBstWouldAddEndPuncttrue
\mciteSetBstMidEndSepPunct{\mcitedefaultmidpunct}
{\mcitedefaultendpunct}{\mcitedefaultseppunct}\relax
\EndOfBibitem
\bibitem[Giamba\c{s}u \latin{et~al.}(2014)Giamba\c{s}u, Luchko, Herschlag,
  York, and Case]{giambasu2014ioncounting}
Giamba\c{s}u,~G.~M.; Luchko,~T.; Herschlag,~D.; York,~D.~M.; Case,~D.~A. Ion
  counting from explicit-solvent simulations and {3D}-{RISM}. \emph{Biophysical
  Journal} \textbf{2014}, \emph{106}, 883--894\relax
\mciteBstWouldAddEndPuncttrue
\mciteSetBstMidEndSepPunct{\mcitedefaultmidpunct}
{\mcitedefaultendpunct}{\mcitedefaultseppunct}\relax
\EndOfBibitem
\bibitem[Giamba\c{s}u \latin{et~al.}(2015)Giamba\c{s}u, Gebala, Panteva,
  Luchko, Case, and York]{giambasu2015competitive}
Giamba\c{s}u,~G.~M.; Gebala,~M.~K.; Panteva,~M.~T.; Luchko,~T.; Case,~D.~A.;
  York,~D.~M. Competitive interaction of monovalent cations with {DNA} from
  {3D}-{RISM}. \emph{Nucleic Acids Research} \textbf{2015}, 8405--8415\relax
\mciteBstWouldAddEndPuncttrue
\mciteSetBstMidEndSepPunct{\mcitedefaultmidpunct}
{\mcitedefaultendpunct}{\mcitedefaultseppunct}\relax
\EndOfBibitem
\bibitem[Nguyen \latin{et~al.}(2019)Nguyen, Yamazaki, Kovalenko, Case, Gilson,
  Kurtzman, and Luchko]{nguyen2019amolecular}
Nguyen,~C.; Yamazaki,~T.; Kovalenko,~A.; Case,~D.~A.; Gilson,~M.~K.;
  Kurtzman,~T.; Luchko,~T. A molecular reconstruction approach to site-based
  {3D}-{RISM} and comparison to {GIST} hydration thermodynamic maps in an
  enzyme active site. \emph{{PloS} One} \textbf{2019}, \emph{14},
  e0219473\relax
\mciteBstWouldAddEndPuncttrue
\mciteSetBstMidEndSepPunct{\mcitedefaultmidpunct}
{\mcitedefaultendpunct}{\mcitedefaultseppunct}\relax
\EndOfBibitem
\bibitem[Naden and Shirts(2015)Naden, and Shirts]{naden2015linearbasis}
Naden,~L.~N.; Shirts,~M.~R. Linear {Basis} {Function} {Approach} to {Efficient}
  {Alchemical} {Free} {Energy} {Calculations}. 2. {Inserting} and {Deleting}
  {Particles} with {Coulombic} {Interactions}. \emph{Journal of Chemical Theory
  and Computation} \textbf{2015}, \emph{11}, 2536--2549\relax
\mciteBstWouldAddEndPuncttrue
\mciteSetBstMidEndSepPunct{\mcitedefaultmidpunct}
{\mcitedefaultendpunct}{\mcitedefaultseppunct}\relax
\EndOfBibitem
\bibitem[Li and Nam(2020)Li, and Nam]{li_repulsive_2020}
Li,~Y.; Nam,~K. Repulsive soft-core potentials for efficient alchemical free
  energy calculations. \emph{Journal of Chemical Theory and Computation}
  \textbf{2020}, \emph{16}, 4776--4789\relax
\mciteBstWouldAddEndPuncttrue
\mciteSetBstMidEndSepPunct{\mcitedefaultmidpunct}
{\mcitedefaultendpunct}{\mcitedefaultseppunct}\relax
\EndOfBibitem
\bibitem[Lee \latin{et~al.}(2020)Lee, Lin, Allen, Lin, Radak, Tao, Tsai,
  Sherman, and York]{lee_improved_2020}
Lee,~T.-S.; Lin,~Z.; Allen,~B.~K.; Lin,~C.; Radak,~B.~K.; Tao,~Y.; Tsai,~H.-C.;
  Sherman,~W.; York,~D.~M. Improved alchemical free energy calculations with
  optimized smoothstep softcore potentials. \emph{Journal of Chemical Theory
  and Computation} \textbf{2020}, \emph{16}, 5512--5525\relax
\mciteBstWouldAddEndPuncttrue
\mciteSetBstMidEndSepPunct{\mcitedefaultmidpunct}
{\mcitedefaultendpunct}{\mcitedefaultseppunct}\relax
\EndOfBibitem
\end{mcitethebibliography}

\end{document}